\documentclass{article}

\usepackage{amsthm}
\usepackage{microtype}
\usepackage{graphicx}
\usepackage{subcaption}
\usepackage{booktabs}
\usepackage{xspace}
\usepackage{amsfonts}
\usepackage{url}
\usepackage[breaklinks]{hyperref} 
\usepackage{xurl} 
\usepackage{hyperref}
\usepackage[accepted]{mlsys2025}
\usepackage{balance}

\usepackage[ruled,vlined,algo2e]{algorithm2e}

\mlsystitlerunning{LAVA: Lifetime-Aware VM Allocation with Learned Distributions and Adaptation to Mispredictions}

\newtheorem{theorem}{Theorem}

\begin{document}

\twocolumn[
\mlsystitle{LAVA: Lifetime-Aware VM Allocation with Learned Distributions and Adaptation to Mispredictions}

\mlsyssetsymbol{equal}{*}

\begin{mlsysauthorlist}
\mlsysauthor{Jianheng Ling}{equal,goo}
\mlsysauthor{Pratik Worah}{equal,goo}
\mlsysauthor{Yawen Wang}{equal,goo}
\mlsysauthor{Yunchuan Kong}{equal,goo}
\mlsysauthor{Anshul Kapoor}{goo}
\mlsysauthor{Chunlei Wang}{goo}
\mlsysauthor{Clifford Stein}{goo}
\mlsysauthor{Diwakar Gupta}{goo}
\mlsysauthor{Jason Behmer}{goo}
\mlsysauthor{Logan A. Bush}{goo}
\mlsysauthor{Prakash Ramanan}{goo}
\mlsysauthor{Rajesh Kumar}{goo}
\mlsysauthor{Thomas Chestna}{goo}
\mlsysauthor{Yajing Liu}{goo}
\mlsysauthor{Ying Liu}{goo}
\mlsysauthor{Ye Zhao}{goo}
\mlsysauthor{Kathryn S. McKinley}{equal,goo}
\mlsysauthor{Meeyoung Park}{equal,goo}
\mlsysauthor{Martin Maas}{equal,gdm}
\end{mlsysauthorlist}

\mlsysaffiliation{goo}{Google}
\mlsysaffiliation{gdm}{Google DeepMind}

\mlsyscorrespondingauthor{Martin Maas}{mmaas@google.com}

\vskip 0.3in

\begin{abstract}
Scheduling virtual machines (VMs) on hosts in cloud data centers dictates efficiency and is an NP-hard problem with incomplete information.  Prior work improved VM scheduling with predicted VM lifetimes.  Our work further improves lifetime-aware scheduling using repredictions with lifetime distributions versus one-shot prediction. Our approach repredicts and adjusts VM and host lifetimes when incorrect predictions emerge. We also present novel approaches for defragmentation and regular system maintenance, which are essential to our data center reliability and optimizations, and are not explored in prior work. We show repredictions deliver a fundamental advance in effectiveness over one-shot prediction.

We call our novel combination of distribution-based lifetime predictions and scheduling algorithms \emph{Lifetime Aware VM Allocation (LAVA)}. LAVA reduces resource stranding and increases the number of empty hosts, which are critical for large VM scheduling, cloud system updates, and reducing dynamic energy consumption. Our approach runs in production within Google's hyperscale cloud data centers, where it improves efficiency by decreasing stranded compute and memory resources by $\sim$3\% and $\sim$2\% respectively. It increases empty hosts by 2.3-9.2 pp in production, reducing dynamic energy consumption, and increasing availability for large VMs and cloud system updates. We also show a reduction in VM migrations for host defragmentation and maintenance.  In addition to our fleet-wide production deployment, we perform simulation studies to characterize the design space and show that our algorithm significantly outperforms the prior state of the art lifetime-based scheduling approach.
\end{abstract}
]

\printAffiliationsAndNotice{\mlsysEqualContribution}

\section{Introduction}
\label{sec:introduction}

Cloud data centers run an increasing fraction of the world's compute. Efficient resource use in data centers is thus critical,  from both an economic and environmental perspective. In Infrastructure-as-a-Service (IaaS) environments, VM scheduling to physical hosts determines a large portion of end-to-end efficiency. Poor assignment of VMs may \emph{strand host resources} – making the remaining host resources too small or too imbalanced to accommodate additional VMs. VM allocation also impacts the number of \emph{empty hosts}, which impacts idle power optimizations, resource utilization, the ability to provision large VMs (obtainability), and system maintenance velocity (e.g., for security patches).

Much prior work addresses workload allocation in data centers \cite{10.1145/2465351.2465386,43438,cortez2017resource,199390,grandl2016graphene,10.14778/2733004.2733012}, including leveraging workload predictions \cite{10.1145/2644865.2541941,10.1145/2451116.2451125,10.1145/3190508.3190515}. The closest related work predicts job lifetimes (time between start and exit) to assist the scheduler in making better placement decisions \cite{yadwadkar2014wrangler,10.1145/2168836.2168847,10.1145/2901318.2901355,cortez2017resource,10.1145/3410220.3456278}. Most recently, Barbalho et al.~\cite{barbalho2023virtual} show bin packing improvements using Lifetime Alignment (LA) scheduling. LA predicts the lifetime of a VM at creation time, and then treats the predicted lifetime as fixed. As a result, mispredictions may tie up an entire host and, over time, model mistakes will accumulate. For example, assume a model with a 1\% chance of predicting a long-lived VM (running for months) as short-lived. If, over time, the scheduler places 70 presumably short-lived VMs on a host, there is a greater than 50\% chance that one of them is long-lived, which would prevent the host from freeing up until that VM exits. Correspondingly, Barbalho et al. show that their best algorithm achieves maximum efficiency with perfect prediction accuracy and degrades rapidly as the accuracy decreases (\cite{barbalho2023virtual}, Figure 8). They thus introduce an alternative algorithm that achieves lower efficiency but is more tolerant to mispredictions.

Our work achieves both tolerance to mispredictions and high efficiency by \emph{repredicting} VM and host lifetimes to correct mispredictions and improve accuracy over time.  We take inspiration from C++ memory allocation work that increases the robustness of lifetime-based allocation by reassessing predictions on the fly \cite{10.1145/3373376.3378525}. For example, if a VM outlives its predicted lifetime, the algorithms change which other VMs are preferred to be co-scheduled with it on the VM's host.  In addition to host bin packing, we address significant challenges not explored by prior work, including host defragmentation, maintenance, and stranding, which are essential components of Google's cloud system design and approach to efficiency. 

Our key contributions are A) VM lifetime reprediction and adaptation to mispredictions, such as when a VM outlives its initial prediction; B) a novel ML model of VM lifetime probability distributions, which captures hard-to-predict VM behavior and prediction uncertainty; and C) three novel lifetime repredicting scheduling algorithms: 1) Non-Invasive Lifetime Aware Scheduling (NILAS) incorporates predictions into an existing scoring function.  2) Lifetime Aware VM Allocation (LAVA) more fundamentally redesigns the scheduler around lifetimes.  3) Lifetime-Aware ReScheduling (LARS) exploits lifetimes to reduce VM disruptions during host defragmentation and maintenance.

We deployed our approach fleet-wide across Google's planet-scale cloud infrastructure. We evaluate it in production and in high-fidelity simulations with production traces, and compare these results to prior work. Some of the key findings of our paper include the following:

\begin{itemize}
\item This paper shows how to exploit our key insight – using probability distribution models of VM lifetimes – to  co-design algorithms that repredict VM lifetimes and actively respond to mispredictions.
\item While NILAS and the prior state-of-the-art (SOTA) LA algorithm place VMs with similar lifetime together, the key idea of LAVA is the opposite – it puts shorter-lived VMs on hosts with one or more longer-lived VM, creating hosts with a wider range of lifetimes. This approach avoids extending the time at which the longest-lived VM will exit and the host will be empty.
\item We show the NILAS, LAVA, and LARS algorithms increase empty host availability (2.3-9.2 percentage points (pp)), reduce stranding ($\sim$2-3\%), and reduce VM disruptions (4.5\%). Note that a consistent 1 pp improvement in stranding or empty hosts can be equivalent to saving 1\% of a cluster's capacity.
\item  Using production data and extensive simulation studies, we show that these algorithms are efficient and robust,  outperforming the prior SOTA.
\item We incorporate lifetime-based scheduling in a complex environment that features dynamic resource management based on VM usage, VM live migrations, and hierarchical scheduling. We show reductions in resource stranding and VM migrations, in addition to improved bin packing quality.
\end{itemize}

Similar to Maas et al.~\cite{10.1145/3373376.3378525}, we find that  on-the-fly adjustment to mispredictions, versus one-off predictions, opens up a fundamentally different class of algorithms and optimizations, in this case for VM scheduling, and perhaps beyond. Our production deployment shows the practicality of this approach, and of deploying ML in the lower layers of the systems infrastructure stack more broadly.

\begin{figure}
    \centering
    \includegraphics[width=\linewidth]{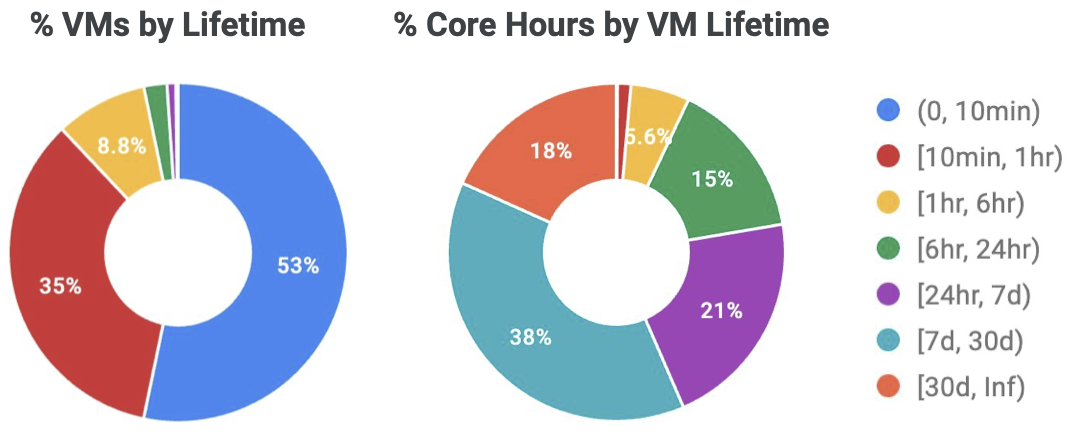}
    \caption{Distribution of VM lifetimes of scheduled VMs vs. their resource consumption.}
    \label{fig:background:lifetime-dist}
\end{figure}
\section{Background}
\label{sec:background}

We first give an overview of VM lifetime properties, our production system, and metrics. Next, we describe the prior SOTA Lifetime Alignment algorithm (LA) \cite{barbalho2023virtual}, to which we compare qualitatively and quantitatively.  

\subsection{Characterizing VM Lifetimes in Production}
\label{sec:background:lifetimes}

In our production environment, most VMs are short-lived, but most core hours are consumed by long-running VMs, similar to the generational hypothesis in garbage collection~\cite{LH:83,Ungar:84}. Figure~\ref{fig:background:lifetime-dist} shows our VM lifetime distributions.  While 88\% of all VMs live for less than 1 hour, 98\% of resources are consumed by VMs that live for 1 hour or more (measured in \emph{CPU cores} $\times$ \emph{time occupied}). Without lifetime-specific handling of VMs, each VM would introduce the same amount of fragmentation, although only the placement of a small fraction of them has an impact on the overall resource efficiency. Prioritizing the bin packing quality of all VMs equally is thus highly inefficient, which motivates lifetime-based VM scheduling.

\begin{figure}
    \centering
    \includegraphics[width=\linewidth]{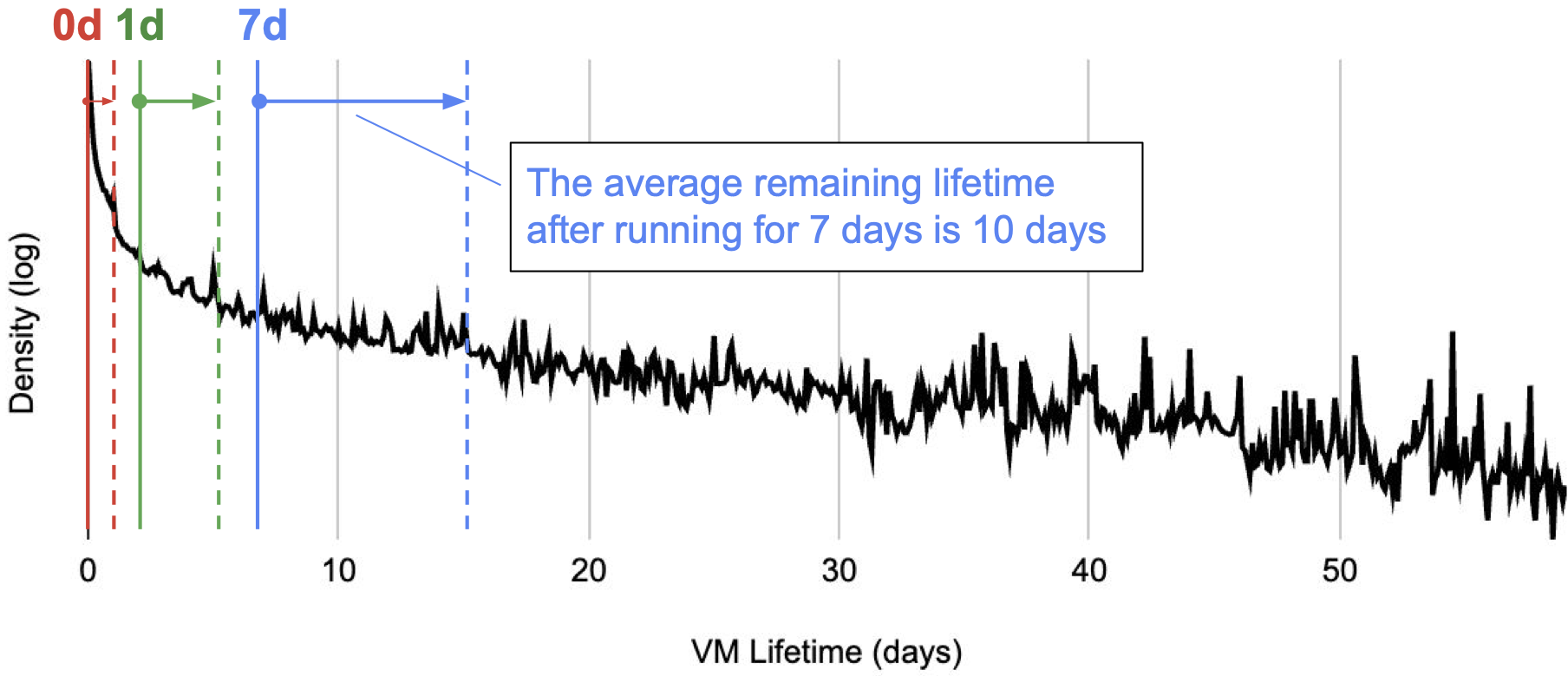}
    \caption{Lifetime distribution (PDF) of VM lifetimes. When the VM is scheduled, the expected (average) lifetime is 0.2 days. After it has run for 1 day, the expected remaining lifetime is 4 days. After 7 days, the expected remaining lifetime is 10 days.}
    \label{fig:background:condprob}
\end{figure}

A key element of our approach is to view VM lifetimes from the perspective of distributions (Cumulative Density Functions or CDFs), as opposed to just averages. We identify that VM features useful for prediction do not fully determine VM lifetimes. The best a traditional model could do would thus be to predict the expected (or average) lifetime for each VM.  Figure \ref{fig:background:condprob} shows the average lifetime may not be particularly meaningful when there are both long and short-lived VMs with the same features. We therefore model and predict the lifetime of VMs as probability distributions. Note that this type of approach has a long history in survival analysis \cite{10.1145/3214306} and in areas such as storage systems \cite{MLSYS2021_efe0df3e}. However, our novelty is predicting distributions and co-designing VM scheduling algorithms to leverage them in detecting and updating mispredictions.

\subsection{VM Allocation}
\label{sec:background:setup}

Our fleet consists of data centers in many geographic regions that run a mix of first and third party workloads. Cloud workloads have distinct host \emph{pools} for distinct VM \emph{families}. VM families represent different products, such as VMs optimized for high performance or cost efficiency. Broadly, VM families fall into two categories. 1) Performance-optimized, slice-of-hardware families assign each VM a fixed partition of a host's resources, e.g., CPU cores, DRAM, SSD, and GPUs. 2) Dynamically sized families share unused resources based on VM and host usage patterns and thus the amount of virtual CPUs and memory of VMs may on occasion exceed the physical resources on a host. This paper focuses on the performance-optimized C2 family and cost-optimized E2 family, representative of these categories.

A scheduling framework called Borg \cite{43438} assigns VMs to hosts. Borg maintains a global view of all VMs and hosts in a given host pool. When a VM creation request arrives, the scheduler computes the set of feasible hosts, i.e., hosts with sufficient resources that match any hard constraints. For each feasible host, it computes a score that determines the preference between them. This score incorporates a number of different business goals, such as spreading tasks across failure domains and bin packing efficiency. This scheduling setup is common \cite{barbalho2023virtual,266915,43438}.

Borg's \emph{Waste Minimization} bin packing algorithm explicitly optimizes for producing empty \emph{shapes} of resources in multiple dimensions (e.g., CPU $\times$ DRAM $\times$ SSD) that match anticipated workload patterns. Borg deployed this approach several years ago because it achieves higher obtainability for large VMs and improves resource efficiency compared to \emph{Best Fit}, the previous scheduler, which was similar to prior work \cite{barbalho2023virtual}. However, Borg's scoring function poses challenges for incorporating VM lifetimes, because it uses lexicographic ordering – one scoring dimension is evaluated at a time, with tie-breakers resolved by the next-lower scoring function. Bin packing is among its final dimensions, ensuring that improving bin packing does not degrade other business objectives.

\subsection{Constraints \& Objectives}
\label{sec:background:constraints}

Our work aims to maximize resource utilization and minimize fragmentation without degrading other business objectives. We measure the following optimization metrics.

\textbf{Empty Hosts.} We measure the percentage of hosts in a pool that is empty. Empty hosts are required to schedule large VMs that consume most or all of the host's resources. They also increase host maintenance velocity when rolling out a kernel or microcode security patch: By increasing empty hosts, applying the update to empty hosts first, and preferring new VMs land on updated hosts, we speed up maintenance and reduce VM disruptions due to live migrations. Additionally, Google puts empty hosts in low power mode and moves empty hosts between pools, configuring them differently to serve various capacity needs. As such, 1 pp of increase in empty hosts directly corresponds to 1\% more capacity available for these use cases. Other metrics of bin packing quality include \emph{packing density} \cite{barbalho2023virtual}, and are equivalent (Appendix~\ref{sec:appendix:bin-packing}).

\textbf{Resource Stranding.} This metric measures free resource shapes and how many future VMs may be scheduled in them. Periodic \emph{inflation} simulations measure stranding.  We take a representative mix of VMs and simulate scheduling as many as possible until capacity is exhausted. The remaining resources on hosts represent \emph{stranded} resources  that cannot fit new VMs. For example, a host may contain free memory but no free CPUs. This memory is thus stranded. The metric is similar to admission control scoring functions~\cite{sajal2023kerveros} and captures unusable capacity when the zone is full.
    
\textbf{VM Live Migrations.} We aim to minimize live migrations. Live migration and fragmentation are closely connected. For some VM families, periodic live migrations will defragment resources. Live migration and resource efficiency may thus be traded against one another. This trade-off is controlled by policy and predates our work. It represents a fundamental difference from some other clouds. Maintenance events trigger live migrations as well.

\subsection{Baseline: Lifetime Alignment (LA)}
\label{sec:background:baseline}

\begin{figure}
    \centering
    \includegraphics[width=\linewidth]{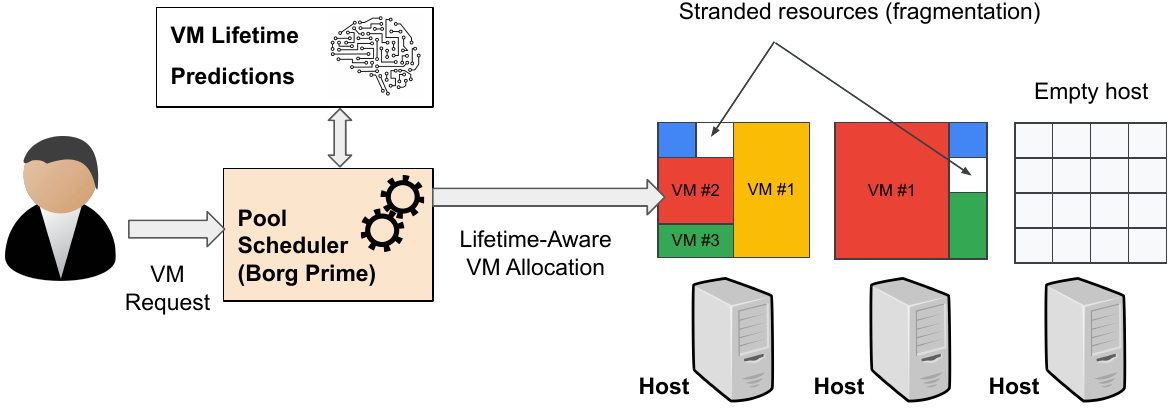}\vspace*{-0.5em}
    \caption{Overview of our VM allocation setup.}
    \label{fig:intro:overview}
\end{figure}

We compare against LA (the \emph{Lifetime Alignment} algorithm) from Barbalho et al. \cite{barbalho2023virtual}. While other work on lifetime-based VM scheduling predates it, LA supersedes this work and, to our knowledge, is the only published approach deployed in production data centers. At a high level, our setup (Figure~\ref{fig:intro:overview}) resembles Barbalho et al.  We use machine learning to predict the lifetimes of VMs and incorporate  predictions into scheduling decisions. Note that their work uses Best Fit scoring. They do not correct mispredictions nor do they consider maintenance and defragmentation, contributions of our paper. 

For every VM allocation request, LA runs a gradient-boosting model to predict its \emph{lifetime class}, which is over an exponentially increasing time interval ($2^i$ to $2^{i+1}$). Their models run on special inference servers. The allocator then preferentially assigns VMs to a host with the same lifetime class, using Best Fit. The lifetime class of a host is defined dynamically: it is the longest \emph{remaining} time of any VM on that host, based on the VM's \emph{initial} lifetime prediction. If there is no host of the same lifetime class, LA picks another suitable host. If no such host exists, it assigns the VM to a new (previously empty) host.

Barbalho et al. report that LA achieves high improvements in packing density with accurate predictions. While LA performs live migrations occasionally, they are not reported or part of the scheduling approach~\cite{barbalho2023virtual}. Because they find the multi-class version of LA is sensitive to mispredictions,  they introduce DPBFR, which uses lifetime predictions only to adjust the quantization of Best Fit scoring. They  deployed DPBFR in production, as it is more robust to mispredictions, but it achieves lower maximum improvement compared to the binary version of LA. We thus compare to a faithful implementation of their best algorithm, called LA-Binary. 

While LA~\cite{barbalho2023virtual} tackled optimizing packing density (empty hosts), we extend the state-of-the-art by reducing stranding and our LARS algorithm decreases the number of migrations needed due to maintenance and defragmentation by determining the migration order. In addition, since we create more empty hosts via lifetime-aware scheduling, we further reduce the need for defragmentation.

\subsection{Lifetime-aware C++ Memory Allocation}
\label{sec:background:llama}

We observe that VM scheduling is similar to explicit memory allocation in C/C++ and other  languages. The memory allocator needs to place objects, that never move, on pages such that pages are either maximally used or become empty \cite{10.1145/3373376.3378525}. VM scheduling is similar. VMs correspond to objects, hosts to pages, stranding to internal fragmentation, and empty hosts to external fragmentation. A key difference is that VM allocation is multi-dimensional.  It manages multiple resources, including CPUs, memory, and SSD,  while memory is one-dimensional (object size).

LLAMA \cite{10.1145/3373376.3378525} is a C++ memory allocator that uses an ML model to predict object lifetimes. Every page is assigned a lifetime class (10ms, 100ms,...) and objects are initially assigned to huge pages with the same lifetime class. Once a page is full, emerging gaps are filled with objects that are \emph{at least one lifetime class lower}. Once all original objects are free, LLAMA reduces the lifetime class of the page by one. This process continues until the page is empty. LLAMA detects mispredictions. If a page is not free before the expected time, LLAMA under-predicted the lifetime of an object, and it increases the lifetime class of the page by one to correct the misprediction. This mechanism minimizes fragmentation while tolerating mispredictions. The latter is critical, since a single misprediction may fragment an entire huge page and    misprediction probability increases exponentially with the number of allocations.

We take inspiration from LLAMA, designing our models and allocators to detect and correct mispredictions.  We cannot directly adopt LLAMA because VM allocation is multi-dimensional and bin packing is not the only business objective. We do, however, incorporate the key insights of LLAMA into new algorithms, NILAS, LAVA, and LARS.
\section{ML Lifetime Prediction Model}
\label{sec:models}

We design ML models that predict the remaining lifetime of a VM both at the time it is scheduled and when re-evaluating the VM and host later, to correct previous mispredictions. Intuitively, reprediction addresses two fundamental issues.

\begin{enumerate}
    \item VM lifetimes depend on many factors that cannot be described by its features, e.g., human behavior or external events. As such, some VMs are fundamentally impossible to predict. Predicting a single VM lifetime thus limits accuracy, as shown in Figure~\ref{fig:background:condprob}.
    \item Estimating the actual lifetime of a previously scheduled VM throughout its execution is not always possible with a single prediction. For example, if a particular VM type's lifetime is bi-modal (e.g., 1 day or 1 week), observing that the VM has been running for 2 days means that it is expected to live for another 5 days.
\end{enumerate}

\noindent Our model addresses these limitations by predicting a probability distribution as opposed to a single value. This approach is well-known in survival analysis \cite{10.1145/3214306}. Survival models  capture the ``uncertainty'' associated with a particular set of feature values. If a VM is highly predictable based on the features, the distribution will be narrow, while difficult-to-predict VMs have wider distributions. Here, we use distributions to update our VM estimates based on the lifetime so far (uptime). Instead of relying on the initial prediction, we compute the \emph{conditional expectation} $\mathbb{E}(T_r | T_u)$ for an updated estimate of a VM's lifetime over time: ``Given a VM has been running for interval $T_u$, what is the expected remaining lifetime $T_r$?''. This value is directly calculated from the PDF in Figure~\ref{fig:background:condprob}.

This approach avails itself to a range of models and we compare a few (Appendix~\ref{sec:appendix:model-comparison}), including different forms of survival models, neural networks, and gradient-boosted decision trees (GBDTs). GBDTs performed best in our experiments. Since they are not distribution-based, we pass an additional parameter $T_u$ into the model that represents the uptime of the VM so far (i.e., the x axis of the PDF), and we predict the remaining lifetime $T_r$. We \emph{augment} every training example by turning it into multiple examples with different values of $T_u$: specifically, 12.5\%, 25\%, ... of the original lifetime. This implementation turns a regression model into a survival model with good accuracy. We train the model directly on $\mathbb{E}(T_r | T_u)$ as the label.

There are two ways to use such a model: 1) When scheduling a VM, run the model repeatedly to materialize a mapping from uptime to the expected remaining lifetime, which may require a large number of model invocations to get sufficient precision. 2) Run the model repeatedly when re-evaluating a VM, which saves resources since it only runs the model for values that are actually needed, but only works if model latency is low. We choose the latter.

Our models use a range of features, including the \emph{zone}, \emph{VM family}, \emph{VM shape}, whether or not the VM attaches to local {SSD}, and internal project metadata (Appendix~\ref{sec:appendix:model-features}). We use a production-grade machine learning library called Yggdrasil~\cite{ydt}. We generate our training data by querying a large-scale internal data warehouse that contains historical data of VMs. We use a combination of distributed SQL queries and data processing pipelines to generate a set of labeled \emph{examples} and feed them into the training framework. Our data sets are on the order of 1M examples for training. The resulting model is loaded from a file system and runs in binaries via a range of language bindings. We periodically retrain our model on recent data.

We train a single joint model for our entire global fleet, across a large number of our data centers. We experimented with training a separate model for each pool, but this approach is disadvantageous from a rollout perspective, more difficult to maintain, and we found that it neither performs better nor significantly improves resource savings. Our model achieves 99\% precision at 70\% recall when used to classify VMs between short and long-lived according to a 7 day threshold. We use this model throughout all of our experiments, to ensure comparability across all algorithms.

An important design consideration is how to deploy the models. For instance, Barbalho et al.~\cite{barbalho2023virtual} run their models on separate inference servers. They cache prediction results to tolerate unavailability of servers, increasing complexity. Running model inference on separate servers also requires handling model rollouts and verification separately since they are not integrated with existing rollout mechanisms. We choose instead to embed the model directly into the Borg binary. This deployment reduces possible failure modes. By piggybacking on Borg rollouts, we update models with the same frequency as Borg, testing the two together. This approach lowers model overhead and achieves a median latency of 9 us (Figure \ref{fig:eval:model_latency}), which is 780$\times$ lower than the median latency for LA \cite{barbalho2023virtual}. With this low latency, we can run our model more frequently, including for lifetime-guided defragmentation (Section \ref{sec:algorithms:defrag}) and to update and correct  mispredictions.

We deployed our model in production over a year ago. It maintained all our production standards, while improving empty machines, stranding, and VM migrations.

\vspace*{1.25mm}
\section{Scheduling Algorithms}
\label{sec:algorithms}

We show that repredicting lifetimes offers a fundamental advantage over one-shot prediction in our multi-dimensional scheduling approach. We introduce three algorithms. NILAS is non-invasive and becomes active when a set of hosts satisfy all other criteria and thus have the same score without the bin packing score (Section \ref{sec:algorithms:las}). LAVA, in contrast, makes lifetime-aware scheduling a fundamental part of the scheduler (Section \ref{sec:algorithms:lava}). Note that while LA and NILAS place VMs with similar lifetime together, the key idea of LAVA is the opposite – it adds many short lived VMs to hosts with one or more long lived VM. The intent is to create hosts that have a wider range of different lifetimes, with an upper bound that is decreasing over time. LARS exploits lifetimes to reduce VM migrations during defragmentation and maintenance (Section \ref{sec:algorithms:defrag}).

\subsection{Theoretical Foundation}
\label{sec:theory}

We show analytically that repredicting lifetimes and correcting for mispredictions fundamentally improves how well any  algorithm can do. Specifically, if the initial error in lifetime prediction is a positive constant, then the number of hosts required without correcting for mispredictions will exceed the number for the same best fit algorithm with repredictions by $\Omega(m)$, where $m$ is the number of hosts. Appendix~\ref{sec:appendix:proof} includes the precise theorem and proof. 

\subsection{NILAS: Non-Invasive Lifetime-Aware Scheduling}
\label{sec:algorithms:las}

\begin{figure}
    \centering
    \includegraphics[width=\linewidth]{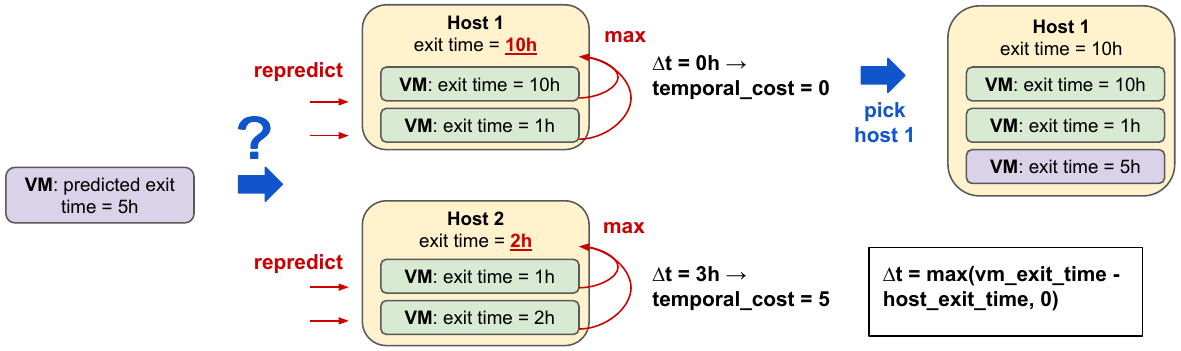}\vspace*{-0.5em}
    \caption{Overview of the NILAS algorithm.}
    \label{fig:algos:las}
\end{figure}

The goal of NILAS (Figure~\ref{fig:algos:las}) is to create more empty machines by scheduling VMs on a host where all other VMs are likely to exit at a similar time or later. Given a VM, NILAS ranks potential hosts  based on the exit times for the VMs already scheduled on that host.  It computes the maximum of the predicted exit time of all VMs on the host. This score is conceptually similar to LA \cite{barbalho2023virtual}, with a crucial difference: Instead of using the original VM lifetime predictions, NILAS repredicts the remaining lifetime (i.e., the \emph{updated} exit time) for all VMs, using their uptime so far.

NILAS corrects any earlier mispredictions as follows. Consider a host where all remaining VMs have exceeded their originally predicted lifetime. Without correcting mispredictions, the scheduler has to assume that this host will soon become empty and will avoid scheduling long-lived VMs on this host. We hypothesize that this choice is the reason why LA is sensitive to prediction accuracy, leading the authors to not deploy LA. In contrast, NILAS schedules new VMs on hosts using the repredicted \emph{remaining} lifetimes of all VMs on each host. As such, NILAS continues to improve its predictions if the original prediction was wrong.

NILAS is integrated into an existing hierarchical scheduler (Section \ref{sec:background:setup}) in a non-invasive way.  When scoring a host, we compute $\Delta T = $ max(predicted\_vm\_exit\_time - host\_exit\_time, 0), where the host\_exit\_time is the maximum of the \emph{repredicted} remaining VM lifetimes on the host. We then quantize this value with bucket boundaries $\{$0m, 30m, 60m, 90m, 2h, 3h, 4h, 6h, 12h, 24h, 168h$\}$. We call the index of the bucket that $\Delta T$ falls into the \emph{temporal cost}. For example, if $\Delta T = 70m$, the temporal cost is 2. This quantization integrates into our lexicographic scoring function (Section~\ref{sec:background:setup}). By quantizing the temporal cost, we create equivalence classes within which hosts can be packed based on additional considerations. This feature integrates well with our multi-dimensional and shape-based bin packing.

We add the temporal cost into our lexicographic scoring function one level above the bin packing score. The temporal cost never has an effect on any higher-ranked business metrics. It only becomes the deciding factor if all those scores are the same. The score is thus  \emph{non-invasive}. It is local to the bin packing score and is equivalent to modifying the bin packing score to add a third dimension (lifetime).

\vspace*{0.8mm}

\subsection{LAVA: Lifetime-Aware VM Allocation}
\label{sec:algorithms:lava}

A limitation of NILAS is that mispredictions may gradually bump up the host lifetime. By trying to match the new VM to the exit times of other VMs on a host, the host may never free up because each under-predicted VM further bumps up the exit time, causing even more VMs to be scheduled, which in turn bumps up the exit time further.

In preference to filling resource gaps on hosts with VMs that exit at a similar time, LAVA attempts to fill gaps with VMs that are at least 10$\times$ shorter lived, so that even mispredictions are unlikely to increase the host lifetime. Further, LAVA updates the host's exit time when detecting a \emph{major} (more than 10$\times$) misprediction. We use order of magnitude differences to limit the impact of small amounts of over or under prediction.  We now describe LAVA in more detail.

LAVA divides lifetime predictions into lifetime classes: $<$1h, 1-10h, 10-100h, 100-1000h. We refer to these lifetime classes as LC1, LC2, L3, and L4. Each host is also assigned a lifetime class. LAVA runs an additional coarse-grained host scoring function based on their lifetime classes and breaks ties with NILAS.

\begin{figure}
    \centering 
    \begin{subfigure}[t]{\linewidth}
        \centering
        \includegraphics[width=\textwidth]{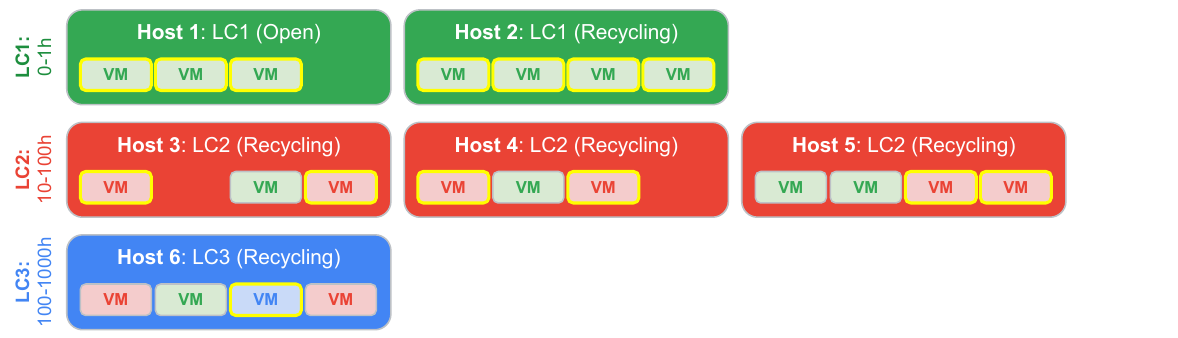}
        \caption{LAVA preferentially places new VMs on \emph{recycling} hosts of a higher lifetime class. If none is available, it chooses \emph{open hosts} of the same lifetime class. Otherwise, it opens a new host.}
        \label{fig:algos:lava1}
    \end{subfigure}
    
    \vspace{0.2cm}
    \begin{subfigure}[t]{\linewidth}
        \centering
        \includegraphics[width=\textwidth]{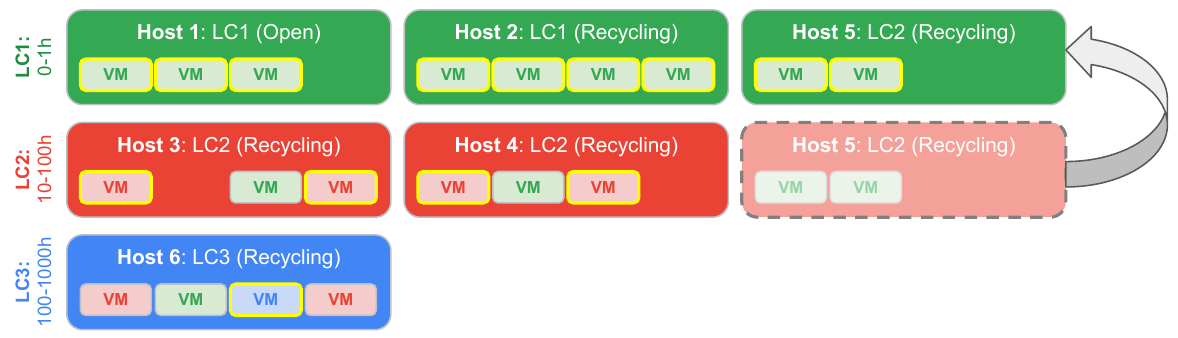}
        \caption{When all residual VMs (highlighted) on a host exit, LAVA reduces the lifetime class by one. All remaining VMs turn residual.}
        \label{fig:algos:lava2}
    \end{subfigure}
    
    \vspace{0.2cm}
    \begin{subfigure}[t]{\linewidth}
        \centering
        \includegraphics[width=\textwidth]{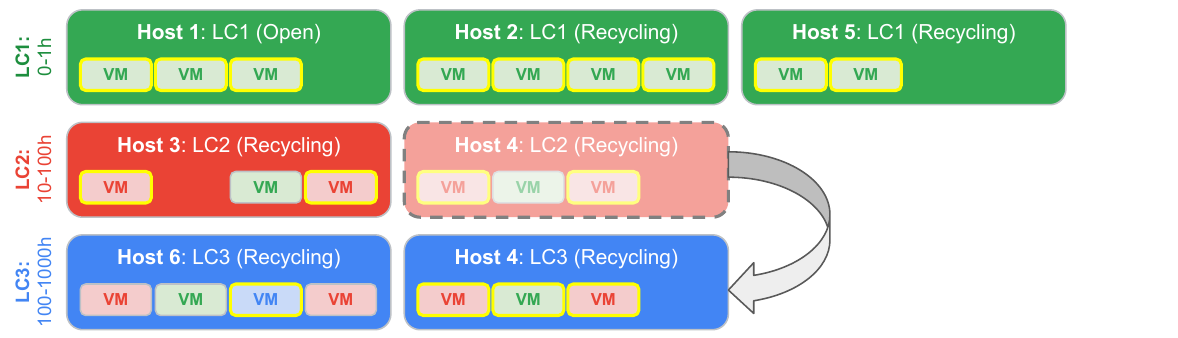}
        \caption{If a host exceeds its deadline, there was a misprediction and LAVA increases its lifetime class by one. Its VMs become residual.}
        \label{fig:algos:lava3}
    \end{subfigure}
    
    \caption{Overview the LAVA Algorithm.}
\end{figure}

Each host has a lifetime class and is in one of three states:  \emph{empty}, \emph{open}, or \emph{recycling}, similar to LLAMA \cite{10.1145/3373376.3378525}. When a new VM arrives and there is no eligible \emph{open} or \emph{recycling} host, LAVA places the VM on an empty host, assigns the host the \emph{open} state, and assigns the host the lifetime class of the VM. As long as this host is open, LAVA will only schedule other VMs of the same lifetime class on it (Figure~\ref{fig:algos:lava1}). Once over 90\% of the resources (CPU or memory) of an open host are occupied, LAVA transitions the host to the \emph{recycling} state. All the VMs that were present on the host when it transitioned state are labeled \emph{residual VMs} (i.e., the VMs of the original lifetime class of the host). \emph{Recycling} hosts  only receive VMs that are at least one lifetime class lower, e.g., LAVA only places new VMs with a predicted lifetime of 10 hours or shorter on 100-hour recycling hosts.

LAVA handles mispredictions as follows. If all predictions are correct, the total lifetime of a host does not exceed 1.1$\times$ its original lifetime class. There are two cases: overpredicted and underpredicted lifetimes.

If lifetimes are overpredicted, VMs will exit early, creating new gaps on the host. The approach of placing shorter-lived VMs in gaps ensures that we keep a host occupied, while minimizing the likelihood that we extend its lifetime. Once all residual VMs exit, we know that all remaining VMs are of the next-shorter lifetime class. We  therefore re-classify the host as one lifetime class lower (Figure~\ref{fig:algos:lava2}), classify all remaining VMs as the new residual VMs, and fill emerging gaps with even shorter-lived VMs. This process repeats until the host is empty.

Conversely, when lifetimes are underpredicted, the host will not become empty within the expected time interval. We  detect this case by establishing a time-out for each host after which we move it to the next-higher lifetime class (Figure~\ref{fig:algos:lava3}). All VMs that are on the host at the time of expiration become the new residual VMs.

LAVA is also integrated into Borg's existing hierarchical scheduler, but adds another scoring function before NILAS. This higher-ranked scoring function keeps track of additional states such as predicted host lifetime class, host state, and deadline for each host. The size of this additional state is negligible compared to other per-VM state. 

When a new VM request arrives, LAVA scores the hosts in the following order (in decreasing preference): \emph{recycling} hosts whose lifetime classes are greater than the VM lifetime class (the closer the more preferred), \emph{open} hosts whose lifetime class are equal to the VM lifetime class, any non-empty host, and lastly empty hosts. In contrast to NILAS, which acts as a tie-breaker for the existing algorithm, LAVA makes decisions that the baseline could not have made, and is therefore \emph{invasive}.

\subsection{LARS: Defragmentation \& Maintenance}
\label{sec:algorithms:defrag}

The above algorithms focus on reducing fragmentation by improving VM placement. Borg also performs software and hardware maintenance activities, applying security, microcode, host OS, and VM software updates.  It also actively defragments hosts for some VM families. These activities live migrate VMs \cite{ruprecht2018vm}.  Live migration uses pre- and post-copy techniques that minimize VM execution time disruptions to seconds. However, total migration times are proportional to VM memory and SSD usage, and consume capacity on both hosts. In simulations, we model that both hosts are busy for a conservative 20 minutes.

Borg reserves a small fraction of hosts for defragmentation, maintenance, and to tolerate hardware failures, and thus these activities do not impact available capacity. We show how using lifetimes can reduce the number of VMs that are disrupted during defragmentation and maintenance. We call the algorithm LARS. The rest of our exposition focuses on defragmentation for conciseness (details in Appendix~\ref{sec:appendix:lars}). 

For some VM families, we defragment hosts when the availability of empty hosts in a particular pool drops below a certain threshold. The defragmenter searches for suitable hosts to defragment, preferring hosts with few VMs and excess resources and to move the VMs to hosts with available slots that match the VM shapes.

LARS first chooses a small number of hosts to defragment based on their excess resources and marks them as unavailable for scheduling.  It then uses predicted lifetimes to determine the order in which to migrate VMs. It starts with the longest-predicted VM and selects a target host to which to migrate it. Choosing this VM reduces the number of migrations, as shorter-lived VMs exit while LARS migrates longer-lived VMs. Once the host is empty, LARS marks it as available to schedule for any lifetime class.

Note that LARS uses the same algorithm to choose a target host for migration as for initial VM placement. As such, NILAS and LAVA use repredictions from our distribution-based model to place the VM on a better-suited host compared to the prior algorithm that was not lifetime-aware.

\vspace*{2.7mm}
\section{Implementation}
\label{sec:implementation}

Our algorithms are integrated into Borg's scheduling code. The scheduler runs within \emph{Borg Prime}, which handles incoming VM requests and assigns them to hosts. The Borg Prime binary is replicated across a number of instances, with one of them serving as the elected leader. All replicas run the same binary and a lower layer implements consensus.

As mentioned in Section~\ref{sec:models}, we compile our model into the Borg Prime binary and roll it out with Borg. This coupling has the advantage that canary deployments (the first step of our rollout process) test the entire end-to-end approach. We deploy our software regularly and could update our model as often as once a week. However, we found that the models' accuracy remains high even a month later as shown in Figure~\ref{fig:eval:model_degradation}. Note that while we found it highly advantageous from a latency and complexity perspective to compile the model into the binary, our approach is not restricted to this deployment scenario.

We carefully measured the CPU and DRAM overheads of our approach in production. Binary size increases by less than 10 MB, and both the memory and CPU consumption of our approach are (generally) indistinguishable from noise. This lack of observable overhead may seem counter-intuitive, but since a VM record has so much other state and Borg performs many other computation during scoring, the extra bytes of memory and the 9 us of model latency are not visible. We process on the order of 10-100 scheduling requests per second in each cluster, and even hypothetically repredicting 100 VMs in the process would add less than 1 ms. In practice, we generally repredict many fewer VMs, since  higher-ranking scoring functions filter out many hosts and we re-score in parallel VMs only on considered hosts.

In some very large zones, re-scoring still became a bottleneck. We therefore added a host lifetime score cache (Appendix~\ref{sec:appendix:caching}). For any host that 1) is assigned a new VM, or 2) an existing VM exits, or 3) the host expected lifetime expires, we re-score the host.  For all other hosts, we re-score them according to a configurable time interval (e.g., 60s). 

\vspace*{1mm}

\subsection{Simulations}
\label{sec:implementation:sim}

\textbf{NILAS/LAVA}: We developed an event-driven simulator to explore our approach.  We extract production traces of VM start, exit, and restart events within a particular pool and then replay this trace against a simulated instance of the scheduler. The simulated scheduler uses our production code base and is thus highly accurate. We validated our simulator against production (Appendix~\ref{sec:appendix:simulator} has more details).

\textbf{LARS}: To calculate the benefit of LARS (lifetime-based defragmentation), we built a modified simulator, using the same production traces. From our traces, we collect the list of migrations that were performed during a time interval. In production, we limit the number of live migrations that can be in progress simultaneously to batches of 3. We simulate this behavior by assuming that all migrations are performed in a certain order (in our baseline, defined by the trace), but have to wait until a \emph{slot} is available. This approach has the effect that some VMs exit while others are migrating. LARS modifies this order based on lifetime predictions.

\subsection{Production Measurements}

Measuring the impact of scheduling policies in production is challenging. Since workloads and total capacity in a pool are constantly changing, it is difficult to attribute  changes in a metric to a particular modification. We therefore use an A/B testing methodology. We split the hosts into two halves and apply our scheduling algorithm to one of them. Our production metrics report empty hosts, stranding, and model prediction for the entire pool. However, stranding is only accurate for entire pools and cannot differentiate A/B hosts. We report live production experiments in four pools, where we apply the algorithm to the entire pool and evaluate scheduling quality. These experiments also ensure that our A/B approach does not distort behavior. Note that these A/B deployments are at production scale. The cluster/zone sizes for this product range from 100s to 10,000+ hosts with O(100-10,000) VMs active at a time.

NILAS has been deployed fleet-wide for one VM family for about a year, at the time of publication. Time series data from the fleet-wide production rollout confirmed our simulation numbers and A/B experiments, and show a fleet-wide increase of empty hosts by 4 pp for this VM family.

\subsection{LA-Binary Comparison}

Our main comparison is to LA-Binary, a faithful implementation of LA, but embedded in the Borg binary versus a separate model service \cite{barbalho2023virtual}. Since our data center environment is different, perfectly replicating their setup is impossible as, e.g., the ML model features are different. We are faithful to LA's description and favorable to LA-Binary when in doubt. For an apples-to-apples comparison, we use the exact same model when comparing LA-Binary to NILAS and LAVA, but without repredictions for LA-Binary. The predictor classifies VMs as short or long lived, using all our model features, but with the same two-hour cutoff threshold as in their paper.

\begin{figure*}
\centering
\includegraphics[width=\linewidth]{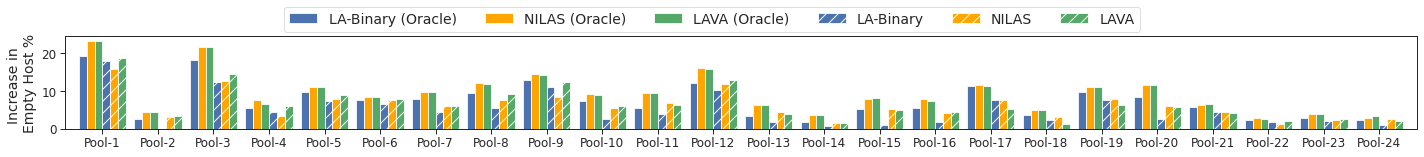}
\caption{Empty host improvements comparing NILAS, LAVA with LA-Binary, to our production baseline for 24 C2 pools in simulation.  On average, LAVA increases empty hosts by 6.5 pp,  compared to 6.1 pp for NILAS, and 5.0 pp for LA-Binary.}
\label{fig:eval:c2}
\end{figure*}
\section{Evaluation}
\label{sec:eval}

This section reports results from production and simulations on production data center traces. Our simulations are highly accurate (Appendix~\ref{sec:appendix:simulator}). We report relative improvements and omit industry-competitive sensitive data.  

\subsection{End-to-end Simulation Results (NILAS \& LAVA)}

We first compare the performance of the different algorithms in simulation using C2 family production traces. Figure~\ref{fig:eval:c2} shows simulation data from C2 traces from 24 pools that cover a wide range of sizes, geographies, and usage patterns. The traces are from May-June 2024 and have a duration of  seven weeks. We measure the average percentage of empty hosts and report changes relative to the production baseline. For instance, if LAVA achieves 21\% and the baseline achieved 20\%, we report a 1 percentage point (pp) improvement. We run all algorithms both with our actual lifetime model and a perfect (oracular) predictor as a comparison point. Similar to the LA paper (which increased packing density by 2\% in production), a consistent 1 pp improvement represents a large gain in this scenario, and can be equivalent to 1\% of a cluster's capacity if used to power down or divest hosts, or offset holdbacks. 

NILAS outperforms LA-Binary on average, achieving an additional 1.1 pp increase in empty hosts. With more optimized misprediction mitigation, LAVA achieves the best average improvement of 1.5 pp compared to LA-Binary. We also observe that NILAS consistently outperforms LA when both have oracular lifetimes: 9.5 pp vs. 7.5 pp on average, confirming our theoretical results that reprediction is fundamentally more powerful than one-shot prediction.

As with many NP-hard problems, we observe occasional inversions when comparing techniques on large numbers of production pools with highly varying workloads. The different production pools we evaluated in Figure~\ref{fig:eval:c2} vary significantly in size, utilization, VM workload distribution, and accuracy of the lifetime model. Isolating the specific factors driving these performance variations is challenging due to the complex interplay of these variables and direct correlations with individual factors were weak at best.

\vspace*{0.8mm}

\subsection{Production Results (NILAS)}
\label{sec:eval:prod}

Prior to fleet-wide rollout, we ran production pilots in waves on C2 and E2 traces. For the first two waves, we conducted A/B experiments in three pools and measured the impact. Table \ref{table:pilot_results} shows our statistically significant empty host improvements,  ranging from 2\% to 10\% for the entire pool. 

\begin{table}[t]
\vspace{-1ex}
\centering
\caption{NILAS Empty Host Improvements in Pilot Pools}
\vspace{1ex}
\small
\label{table:pilot_results}
\begin{tabular}{@{}l c l@{}}
\textbf{Pilot Pool} & \textbf{Type} & \textbf{Change in Empty Hosts}\\ \toprule
C2 Wave 1 pool       & A/B             & +2.3 pp (p-value = 0.01)                                                                \\ 
C2 Wave 2 pool 1       & A/B             & +2.7 pp (p-value $<$ 0.01)                                                                \\ 
C2 Wave 2 pool 2       & A/B             & +9.2 pp (p-value $<$ 0.01)                                                                \\ 
C2 Wave 3 pool       & All      & +4.9 pp (95\% CI: [0.54, 9.2])     
                                          \\ 
E2 Wave 1 pool       & All      & +6.1 pp (95\% CI: [1.9, 10.0])     \\ \bottomrule
\end{tabular}
\caption{VM Migration Reductions Using LARS on Two Traces.}
\vspace{1ex}
\centering
\small
\begin{tabular}{@{}l r| r r r@{}}
& & \multicolumn{3}{c}{\textbf{Migrations}} \\
& \textbf{Scheduled} & \textbf{Baseline} & \textbf{LARS} & \textbf{Reduction} \\
\toprule
1 & 48,239 & 37,108 & 35,505 & 4.32\% \\
2 & 53,597 & 36,307 & 34,655 & 4.55\% \\
\bottomrule
\end{tabular}
\label{fig:eval:defrag}
\end{table}

\begin{figure}[t]
    \centering
    \begin{subfigure}{\linewidth}
    \includegraphics[width=\linewidth]{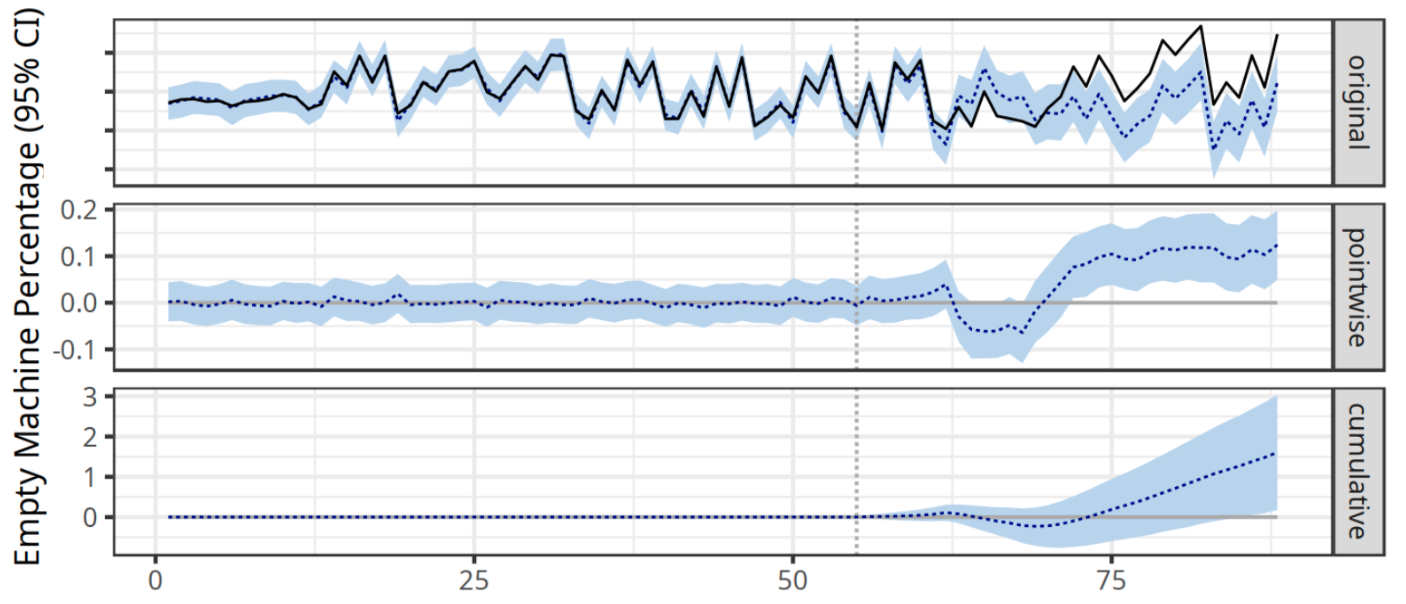}
    \caption{Empty hosts.}
    \end{subfigure}
    \begin{subfigure}{\linewidth}
    \includegraphics[width=\linewidth]{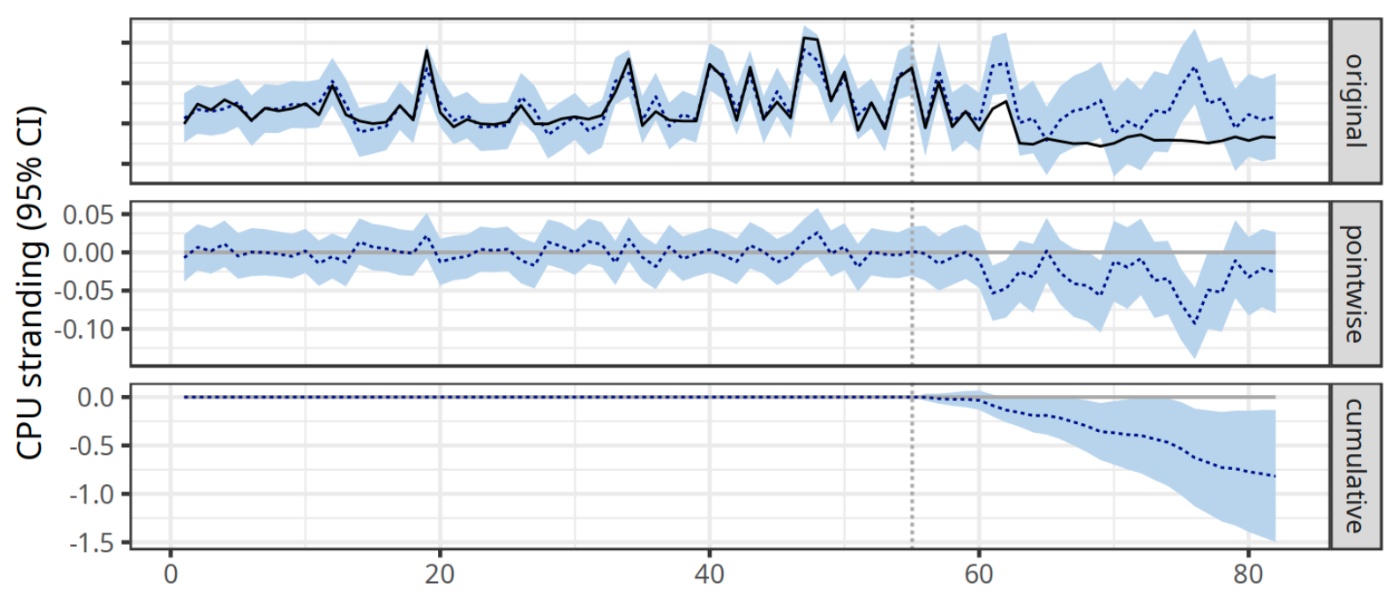}
    \caption{CPU stranding.}
    \end{subfigure}
    \caption{Time series from our Wave 3 rollout, produced by CausalImpact \citep{CausalImpact}. Each sub-figure has three panels. The first panel shows empty machines and a counterfactual prediction (dotted line)  post-launch (labels omitted due to sensitivity). The second panel shows the difference between observed data and counterfactual predictions, i.e., the point-wise causal effect, as estimated by the model. The third panel adds the point-wise contributions from the second panel, plotting the cumulative effect.}
    \label{fig:eval:wave3}
\end{figure}

For the third wave, we piloted NILAS on the entire pool. We employ a Bayesian Structural Time Series method CausalImpact \citep{CausalImpact} to evaluate the pre-post causal effect of NILAS (Figure~\ref{fig:eval:wave3}). We observed a significant empty host improvement of 4.9\% (Table \ref{table:pilot_results}). Moreover, for the whole-pool pilot, we additionally measure the causal effect of NILAS on stranded CPU and memory, where 1 pp of stranding reduction translates directly into 1\% of capacity. NILAS reduced CPU stranding by 3\% (95\% CI: [-5.5\%, -0.49\%]) and memory stranding by 2\% (95\% CI: [-3.8\%, -0.23\%]). We also conducted a whole-pool pilot for an E2 pool. We note that C2 is slice-of-hardware while E2 is dynamically sized. The fact that the approach worked for both types of VM families is a non-obvious result, since dynamic resizing of  VMs further complicates bin packing.

\subsection{Lifetime-based Host Defragmentation (LARS)}

Using production traces, we simulate defragmentation on two different one-month time intervals from a number of pools with customer workloads, including C2 and E2 VMs. In this experiment, we use oracle lifetimes. LARS does not affect which hosts are picked for migration at a given time, but orders the VMs on each in-migration host so that the VMs with the longest remaining lifetime are migrated first (Appendix~\ref{sec:appendix:lars}). The effect is that a larger number of VMs terminate before they would have been migrated, resulting in a reduction of around 4.5\% of live migrations.

\subsection{ML Model}
\label{sec:eval:model}

Our lifetime model achieves 70\% recall at 99\% precision, which is significantly higher than other model types we tried (Appendix~\ref{sec:appendix:model-comparison}). We measured the latency of our model and find that most predictions complete in under 10 us (Figure~\ref{fig:eval:model_latency}). This latency improves upon \cite{barbalho2023virtual} by 780$\times$, and shows the benefit of compiling the model into the Borg Prime binary. Fast predictions make it practical to frequently re-run the model to correct mispredictions and to use it during maintenance and defragmentation.

We analyze the impact of model features in Appendix~\ref{sec:appendix:model-features}. We also analyze the sensitivity of NILAS and LAVA to prediction accuracy and show that our improvements apply across accuracies (Appendix ~\ref{sec:appendix:em-tradeoff}). Finally, we performed a number of ablations to show that NILAS with oracle lifetimes and positioned high in the scoring function achieves close to the theoretical maximum of empty hosts (Appendix~\ref{sec:appendix:ablations}). The gap between production performance and theoretical optimum is a combination of prediction accuracy and priority within the scoring function.

\subsection{Impact of Repredictions}

\begin{figure}
    \centering
    \includegraphics[width=0.95\linewidth]{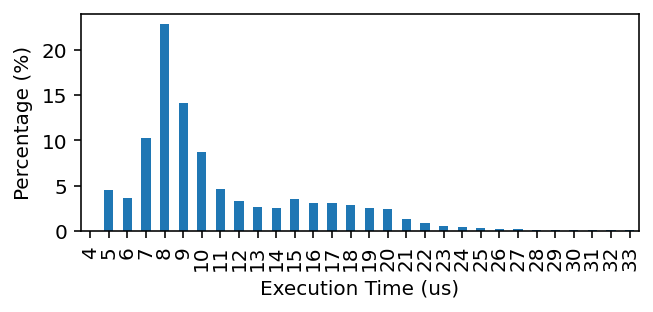}
    \vspace*{-1.5em}
    \caption{Histogram of model execution latencies.}
    \label{fig:eval:model_latency}
    \centering
    \includegraphics[width=0.95\linewidth]{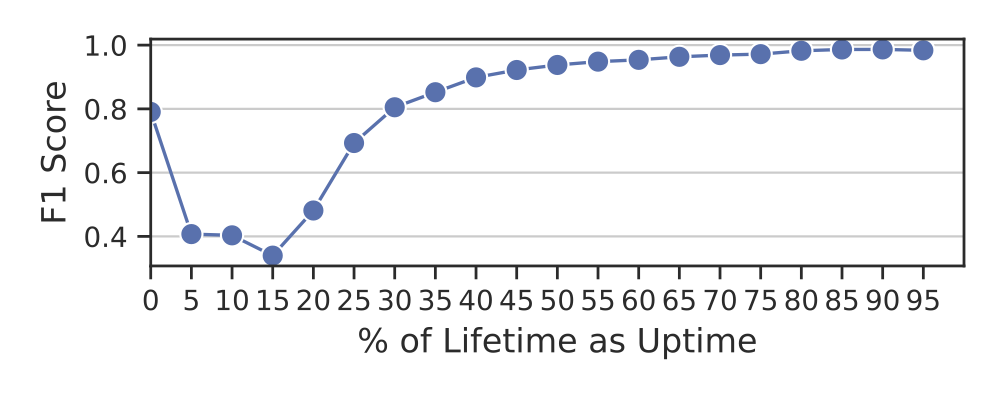}
    \vspace*{-2em}
    \caption{Making repredictions of VM remaining lifetime using its uptime significantly improves the model accuracy.}
    \label{fig:eval:uptime_accuracy}
\end{figure}

To further understand the utility of repredictions based on the current VM uptime, we vary the amount of uptime used as input feature to the model and assess its impact on model accuracy. For each VM in the test trace, we divide its total lifetime into 20 quantiles. We then show the F1 score for predicting whether or not a VM lives beyond 168 hours (7 days) when the uptime is set to a particular quantile. Figure~\ref{fig:eval:uptime_accuracy} illustrates the relationship between varying amount of uptime and model accuracy. We observe that without using reprediction (no uptime/0$^{th}$ quantile), the model only has a F1 score of 0.8. The F1 score quickly rises above 0.9 after the 8$^{th}$ quantile (i.e., once the VM's uptime has reached 40\% of its total lifetime). Observing how long a VM has lived and repredicting its remaining lifetime thus leads to more accurate predictions. We find repredicting longer-lived VMs is critical to avoid spreading them across hosts and causing fragmentation.

One surprising result is that the the F1 score of quantiles 1-5 is, in fact, lower than the accuracy at the 0$^{th}$ quantile. This is because these uptimes are very close to 0 in absolute terms and thus difficult for the model to disambiguate, particularly since we are operating in the log domain. In contrast, the accuracy at 0\% uptime is higher because it is always 0 and the model has seen many training samples with uptime=0. A potential optimization would be to only pass the uptime to the model if it reaches a particular threshold (e.g., 30 seconds), to avoid this initial dip in accuracy.

\subsection{Decrease in Model Accuracy Over Time}

\begin{figure}
    \centering
    \includegraphics[width=\linewidth]{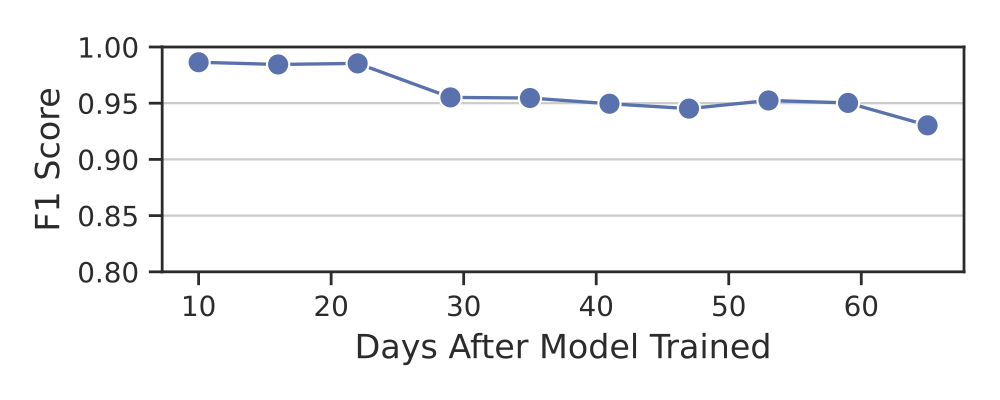}
    \vspace*{-2.5em}
    \caption{The model remains high accuracy weeks after training, making its maintenance overhead low for production deployment.}
    \label{fig:eval:model_degradation}
\end{figure}

We find that the accuracy of our model remains stable for several months after deployment (Figure~\ref{fig:eval:model_degradation}), but does require periodic retraining on the order of once a month to maintain its high accuracy. There are at least two reasons why accuracy drops over time. 1)  New workloads arrive that the model was not trained on. 2) Workloads change their behavior. We looked at production data during this time period and observed gradual shifts in resource usage that is consistent with  shifting workloads.
\section{Production Experience}

NILAS has been running in production since early 2024. We now share general insights from our experience of integrating ML into a mature production system. Our project started in 2020 and we quickly developed initial versions of our VM lifetime models and algorithms in simulation. Most time was spent on putting the approach in production.

\paragraph{Explainable models} One key requirement was that models had to be interpretable and explainable to be accepted in a production environment. We started with a lookup table approach where each entry contained a survival curve produced using Kaplan Meier \cite{Kaplan-Meier}. However, we found that the resulting models underperformed (Appendix~\ref{sec:appendix:model-comparison}) and were not space-efficient. GBDTs showed the best combination of accuracy and explainability.

\paragraph{Model Deployment} We initially considered running the model within separate inference servers, but switched to an in-binary approach. Running our models at $<$10us latency became critical for our reprediction-based approach. We also found that depending on separate servers would have added a circular dependency from a reliability standpoint, since those servers would themselves run on top of Borg. For the same reason, we included the model in the binary as opposed to loading it from, e.g., a distributed file system.

\paragraph{Rollout} A major concern in our production environment was how to roll out the model. Rolling out the model independently of Borg risked the model becoming a single point of failure by bypassing existing verification approaches (e.g., gradual rollout and canary deployments). We found that treating the model like any other code change and rolling it out with new versions of the binary side-stepped this issue by subjecting models to the same careful production testing. We also relied extensively on simulations for backtesting and tuning design parameters, to avoid production changes.

\paragraph{Production Monitoring} During production rollout, we carefully monitored the behavior of NILAS, by adding additional telemetry and monitoring dashboards. This telemetry captures, e.g., the increase in CPU and memory caused by loading the model in production, the increase in time each scheduling pass takes, and how often the new cost component becomes the tie breaker. This instrumentation confirms the system is working as intended and debugs production issues. For example, in one of our initial deployments, we found a bug where the model was called incorrectly.
\section{Related Work}
\label{sec:related}

Prediction-based job scheduling has a long history. Some work uses reinforcement learning \cite{10.1145/3341302.3342080,10.1145/3472883.3486971,10.1145/3503222.3507704}, e.g., for placing containers. Other work predicts properties of jobs and uses them to improve scheduling for scale-out \cite{10.1145/2541940.2541941}, interference \cite{10.1145/2451116.2451125}, or goodput~\cite{10.1145/3600006.3613175}.

Job lifetimes are particularly useful for scheduling. Jockey \cite{10.1145/2168836.2168847} and Aria \cite{10.1145/1998582.1998637} predict lifetimes (called runtimes) in data processing frameworks and use them to improve scheduling of dataflow-based jobs. TetriSched \cite{10.1145/2901318.2901355} uses runtime predictions in conjunction with an MILP solver to schedule jobs in an HPC-style setting. Wrangler predicts stragglers \cite{yadwadkar2014wrangler}. 3Sigma \cite{10.1145/3190508.3190515} predicts various properties of Google job traces and uses them for scheduling. Note that most of this work uses runtime predictions in domain-specific scenarios and does not leverage general machine learning techniques to schedule. 

In Cloud Computing, Microsoft's Resource Central system \cite{cortez2017resource} performs predictions on VMs and uses them for various tasks, including VM scheduling and admission control \cite{sajal2023kerveros}. Microsoft's LA algorithm that performs lifetime-based VM scheduling is most similar to our work \cite{barbalho2023virtual,10.1145/3410220.3456278}. Our work differs in that instead of performing one-shot predictions when the VM starts running, it repredicts VM lifetimes as needed, updating lifetimes based on new knowledge, delivering good improvements in empty hosts over LA, as shown in Section~\ref{fig:eval:c2}. We show how to incorporate lifetimes in our existing scheduler by integrating the models into the binary. Our  models run in-process at very low latency and our new scheduling algorithms tolerate mispredictions. LAVA changes the objective function from LA and NILAS -- instead of placing VMs with similar lifetimes on the same host, LAVA seeks to make more hosts empty by \emph{not} extending the host lifetime, by placing short lived VMs on a host with one or more long lived VM(s). 
\section{Conclusion}
\label{sec:conclusion}

We presented a new approach to lifetime-aware VM allocation in cloud data centers. Our approach continually repredicts VM and host lifetimes, and corrects for past mistakes. NILAS is deployed fleet-wide for one VM family at Google, where it shows significant improvements by increasing empty hosts and reducing stranding.

\vfill
{\fontsize{8}{4}%
\linespread{0.5}
\selectfont\textbf{Acknowledgements:} We would like to thank Deniz Altınbüken, Mike Dahlin, Vahab Mirrokni, Andreas Terzis and the anonymous reviewers for their feedback. We thank Wan Chen, Eileen Feng, Vijayan Satyamoorthy, Dmitry Shiraev, Samuel Smith and Sirui Sun for their contributions.

}

\newpage

\balance

\clearpage

\nobalance

\appendix
The following appendices explore our lifetime model, algorithmic configurations, metrics,  include a proof of correctness,  provide additional details on simulations, and present ablation studies.

\section{Model Features}
\label{sec:appendix:model-features}

Our model utilizes a set of features specific to our cloud environment. Table \ref{fig:detailed-features} shows the list of  features and a brief description of each.

Some of the categorical features, e.g., Metadata ID, VM Category, and VM Shape, take on a large range of different values. We reduce this number by collapsing any category with less than 10 examples in the training set  to a catch-all ``Other'' category. This categorization avoids overfitting and makes the number of different categories manageable.

\begin{figure}
    \centering
    \includegraphics[width=\linewidth]{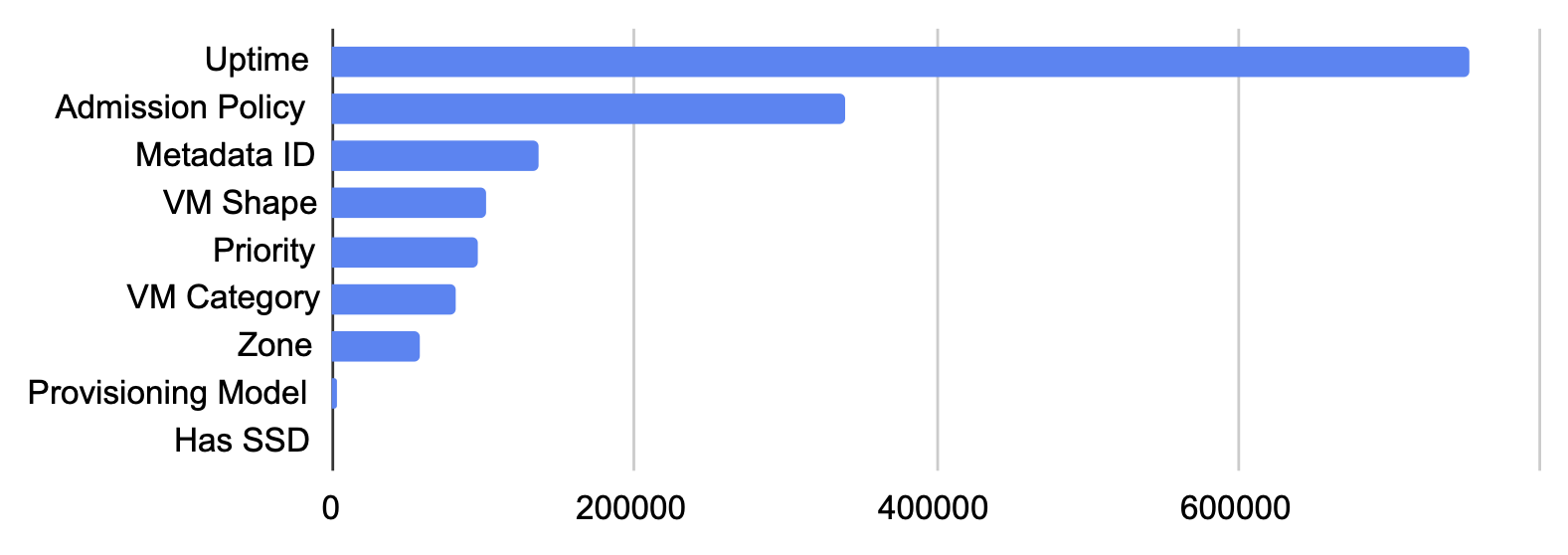}
    \caption{Impact of different features of our model on the prediction accuracy, based on the \emph{split score} of each feature.}
    \label{fig:eval:feature_importance}
\end{figure}

We analyzed the importance of different features of our model (Figure~\ref{fig:eval:feature_importance}), using our decision forest library's \emph{split score}, which is an indication of how much a given feature influences the model's decision. We found that the admission policy plays a major role, which identifies certain special VMs that are admitted without a quota check. We also found that the host pool and the shape of the VM play an important role in prediction.

\begin{table*}[th]
\centering
\caption{Description of model features.}
\vspace{1ex}
\label{fig:detailed-features}
\begin{tabular}{l|l|l} 
\textbf{Feature} & \textbf{Type and Cardinality} & \textbf{Description} \\
\hline
Zone & Categorical (High) & The geographical zone the VM is running in. \\
VM Shape & Categorical (High) & The resource dimensions associated with the VM. \\
VM Category & Categorical (High) & A tag indicating an internal VM categorization. \\ 
Metadata ID & Categorical (High) & An internal ID to group certain related VMs together. \\
Has SSD & Boolean (Low) & Does the VM have SSDs associated with it? \\
Provisioning Model & Boolean (Low) & Is the VM a spot instance or on-demand? \\
Priority & Categorical (High) & Some VMs can be pre-empted and have lower priority. \\
Admission Policy & Boolean (High) & Whether to admit without a quota check (used in special cases). \\ 
Uptime & Float (High) & The uptime of the VM so far, in hours (log). \\
\end{tabular}
\end{table*}

\begin{table*}
    \centering
    \caption{Comparison of different lifetime models.}
    \vspace{1ex}
    \small
    \begin{tabular}{@{}l l l r r r r@{}}
         \textbf{\textsf{Model}}  & \textbf{\textsf{Type}}  & \textbf{\textsf{Library}}  & \textbf{\textsf{C-index}}  & \textbf{\textsf{Precision}}  & \textbf{\textsf{Recall}}  & \textbf{\textsf{F1 Score}} \\ \toprule
            Linear Cox  & survival     & Sksurv    & 0.52 & 0.97 & 0.64 & 0.77\\
            Stratified KM  & survival   & Lifelines & 0.73 & 0.38 & 0.38 & 0.38\\
            Tree-based Cox  & survival & Xgboost   & 0.78 & N/A & N/A & N/A \\ 
            Neural Network & regression& Keras     & 0.73 & 0.99 & 0.58 & 0.73 \\
            \textbf{Gradient-Boosted Decision Trees (GBDT)} & \textbf{regression}    & \textbf{Yggdrasil}     & \textbf{0.84} & \textbf{0.99} & \textbf{0.70} & \textbf{0.8} \\ \bottomrule 
    \end{tabular}
    \label{tab:models:accuracy}
\end{table*}

\vspace*{1.3mm}

\section{Model Baselines}
\label{sec:appendix:model-comparison}

We compare a range of different model types and libraries to identify the ideal model for our use case. We considered both survival models and regular regression models, with model structures varying across linear, tree-based and neural network-based. Among the libraries that we experiment with, Sksurv \citep{sksurv} is a standard Python library for survival analysis supporting the standard linear Cox model \citep{Cox}; Lifelines \citep{lifelines} is a similar Python survival library allowing us to compare with a stratified Kaplan-Meier survival model \citep{Kaplan-Meier}; Xgboost \citep{xgboost} provides implementation of a non-linear Cox model using tree ensembles; we employ Keras \citep{keras} and Yggdrasil Decision Forests \citep{ydt} for a standard regular neural network regression and a tree-based ensemble regression, respectively. We also experimented with different hyperparameters.

Table~\ref{tab:models:accuracy} compares models and shows that the gradient-boosted decision trees (GBDT) models perform best, achieving 99\% precision at 70\% recall when used to classify VMs between short and long-lived according to a 7 day threshold. We thus use these models in this paper and in production.

We experimented with regression models treating lifetimes both as linear and Log10. We found the latter to work better, since lifetimes span many orders of magnitude, which causes difficulties for regression models. We thus use Log10 for all lifetimes in our final model, including VM uptime.

For our production model deployments, we found that the accuracy of the model increases if we cap VM lifetimes at 168 hours (7 days). In production, all VMs with a lifetime longer than 7 days are capped.  We found that capping avoids the case that a small number of very long-lived VMs ``distract'' the model.

\begin{figure}
    \centering
    \includegraphics[width=\linewidth]{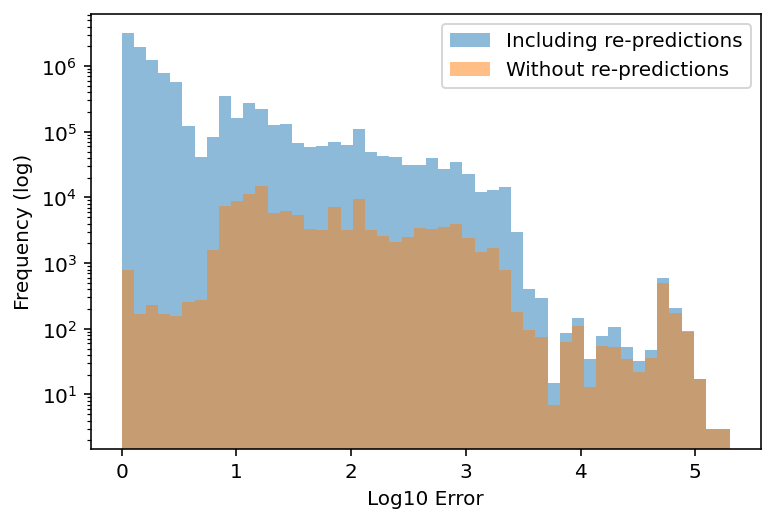}
    \caption{Histogram of the error of our GBDT model within the Log10 domain, for the first 10 million unique predictions of running NILAS with a trace.} 
    \label{fig:eval:error_dist}
\end{figure}

For our final Yggdrasil GBDT model \cite{ydt}, we use all the defaults, except for the following hyperparameters:

\begin{itemize}
    \item Number of Trees = 2000
    \item Maximum Number of Nodes = 32
    \item Growing Strategy = Best First Global
\end{itemize}

\section{Model Prediction Accuracy}
\label{sec:appendix:model-accuracy}

To better understand the performance of our model, we added instrumentation to our simulation runs that log every invocation of the model and the associated prediction. Since our traces contain the ground truth, we can then compare the prediction to the actual lifetime of the VM.

We ran with NILAS against one of our traces and recorded the first 10M predictions. Figure~\ref{fig:eval:error_dist} shows the error distribution, both with and without including repredictions of the same VM. Note that both the x and y axis are log scale (i.e., a much larger fraction of the errors are within a factor of 10 – or 1 in the Log10 domain – than it might appear from the graph). Given a prediction $\hat{y}$ and ground truth $y$, both in seconds, we calculate the error as follows:

\begin{math}
\textit{Error} = | \log_{10}(\hat{y}) - \log_{10}(y) |
\end{math}

One surprising result of this experiment was that the error distribution is not Gaussian or bimodal. Instead, we see a distribution with multiple peaks. We also observe that the distribution that includes repredictions skews significantly more to the left (i.e., lower error) than the distribution without, which confirms that repredictions improve accuracy.

\section{Comparing Bin Packing Metrics}
\label{sec:appendix:bin-packing}

Throughout the paper, we use \emph{empty host percentage} as the main metric to capture bin packing quality. Empty host percentage is the fraction of hosts in a pool that are fully empty. The reason for this choice is that empty hosts most directly map to efficiency gains. Improving the empty host percentage by one percentage point (pp) corresponds to 1\% of the pool's capacity that is now available to put in lower power mode, divest, or power down completely. Alternatively, this 1\% is available to schedule large VMs that otherwise could not have been scheduled due to fragmentation. Finally, as we state in Section~\ref{sec:algorithms:defrag}, there is always a small fraction of hosts that are  required for maintenance and other purposes, and empty hosts make those processes faster.

Since other papers use equivalent but slightly different metrics to measure bin packing quality, we want to briefly highlight how these metrics compare:

\textbf{Empty-to-Free Ratio}: This alternative metric measures empty hosts. It is defined as the fraction of free CPU resources that are in completely empty hosts, containing no VMs (i.e., the number of CPU cores on empty hosts divided by the total number of free CPU cores). This metric has the advantage that it is independent of the utilization of the pool and thus helps normalize changes across pools. On the other hand, the magnitude of the resulting changes is less meaningful and we do not report them in the paper. When comparing different algorithms on the same trace, empty-to-free ratio and empty hosts behave the same, since utilization and pool size are identical between all runs.

\textbf{Packing Density}: Packing density is used by Barbalho et al. \cite{barbalho2023virtual}. It is defined as the number of allocated cores on non-empty hosts divided by the total number of cores on non-empty hosts.

\begin{figure}
    \centering
    \includegraphics[width=\linewidth]{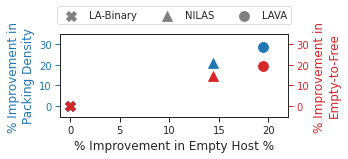}
    \caption{Comparison of different metrics from simulations done on one pool, showing relative improvements from LA-Binary.} 
    \label{fig:eval:metric}
\end{figure}

In our setup, the three metrics are correlated as shown in Figure~\ref{fig:eval:metric}, since all hosts within a pool have the same number of CPUs and the composition of the pool does not change throughout the experiment. As such, improving one results in a corresponding improvement of the others.

\vspace*{2mm}

\section{Analytical Proof}
\label{sec:appendix:proof}

We prove the theorem introduced in Section~\ref{sec:theory} that shows an algorithm that repredicts VM lifetimes fundamentally outperforms an algorithm that does not.

To capture the  benefit of correcting mispredictions, we consider a simplified model in which there are only two job lifetimes, which we call short ($S$) and long ($L$).  We also use $S$ and $L$ to denote the lifetimes themselves.   When a job $j$ arrives, it has a predicted lifetime and a real lifetime.  We know the predicted lifetime and we only learn the real lifetime by running the job.    An algorithm  assigns it to a host  using some combination of a packing score and the lifetime prediction. For simplicity, we  consider a best fit criteria combined with predicted lifetimes.  Assuming that $S << L$ and the jobs arrive at a rate of $\lambda \geq 1/S$, the algorithm will place a job on a host with other jobs with the same predicted lifetime, thereby keeping a small number of hosts partially filled at any time.  Although job sizes and the number of jobs per machine is variable, for simplicity,  assume  each of $m$ hosts can hold at most $k$ jobs.   

If the predicted lifetimes are correct, this algorithm will tend to partition the jobs based on lifetime, which is basically an optimal algorithm under any reasonable set of conditions.  Predictions are not perfect, but  we can learn over time.   We assume an error rate of $\epsilon$, i.e., an $\epsilon$ fraction of the jobs'  original lifetimes are not correct.  Given two job lifetimes, once a job has run for $S$ units of time, we learn whether it is short or long.   If the algorithm then reschedules the job, we have transformed it to a perfect predictor, so we will not allow reassignment, and study  the impact of having this additional knowledge.  

Observe that the errors are asymmetric. If we predict that a job is long when it is  short  then we can recover quickly by assigning other short jobs to that machine.  On the other hand, if we predict that a job is short when it is long, we now have a long job on this machine that otherwise presumably had short jobs.  This case is exactly where continuous learning  helps us.  We classify a host as $L$ if it has any jobs whose processing time is known to be $L$ and $S$ otherwise.   Our algorithm will put new predicted $L$ jobs on $L$ hosts and new predicted $S$ jobs on $S$ hosts.

Now consider the effects of learning (i.e., repredictions) versus not learning. We want to compute, in an interval of time of length $x$, the probability that an $S$ host has both $L$ and $S$ jobs.   The probability that one job is mispredicted is $\epsilon$, and in an interval of length $x$, we will have, in expectation,  $\rho \lambda x$ $L$ jobs, where $\rho$ is the fraction of jobs that are $L$ jobs.   Thus the probability that there is at least one error is at least
\begin{equation}
\label{eq:pr}
1 - (1 - \epsilon)^{(\rho \lambda x)} \ . 
\end{equation}
Thus, if $x \leq 1/(\epsilon \rho \lambda)$, and let let $V$ be the event that there is an error in an interval of length $X$, we have
\begin{eqnarray*}
\Pr[V] & \geq & 
1 - (1 - \epsilon)^{(\rho \lambda x)} \\
& \geq & 1- (1-\epsilon)^{ \rho \lambda/(\epsilon \rho \lambda)} \\& = &  1- (1-\epsilon)^{1/ \epsilon} 
\\ & \geq & 1- 1/e \ ,
\end{eqnarray*}
where the last inequality uses the bound $(1- 1/z)^z \geq 1/e$, with $z=1/\epsilon$.

Next, consider the case where we can learn the true values of jobs' lifetimes after they have been running for $S$ units of time.  We will show that we  can now tolerate more errors, because if we learn that a host has both $L$ and $S$ jobs on it, we just reclassify it as a $L$ host and continue. As the $S$ jobs exit, we put more $L$ jobs on the now correctly classified $L$ host. The misclassification error does make the host unavailable for future $S$ jobs but we still have all the other machines on which to put $S$ jobs.  We only have a problem, when we have many too many $L$ machines and therefore, no room for the $S$ jobs.  In order to get to this position, we  we will need to have  $cm$ such misclassified jobs, which will remove a constant fraction of the capacity and impact the ability to schedule $S$ jobs.

We now show that even in an interval that is $\Omega(m)$ longer than in the previous cases, we will not have enough errors to make $S$ jobs unscheduleable.  
Consider an interval whose length is $cmx/2$, which is $\Theta(m)$ longer than the interval in the no learning case.   The expected number of misclassifications in such an interval is now $\epsilon \rho \lambda cm x/2 = cm/2$. But, as mentioned above, we can now tolerate $cm$ errors.   Applying standard Chernoff bounds, we see  that the probability that we get $cm$ errors (twice the expectation of $cm/2$) is small.

The above discussion for our simplified model can be condensed into the following statement:
\begin{theorem}
Suppose that job lifetimes are either Short or Long, and that jobs are selected independently, to arrive at a constant rate, to be scheduled on one of $m$ hosts, each of constant capacity. If the initial error in lifetime prediction is a positive constant, then the number of hosts required in the best fit scheduling algorithm without learning will exceed the same best fit algorithm with learning by $\Omega(m)$. 
\end{theorem}

\section{Simulator Details \& Validation}
\label{sec:appendix:simulator}

We provide additional details about our time series-based simulator that we use for evaluating NILAS and LAVA.

\begin{figure}
    \centering
    \includegraphics[width=0.45\textwidth]{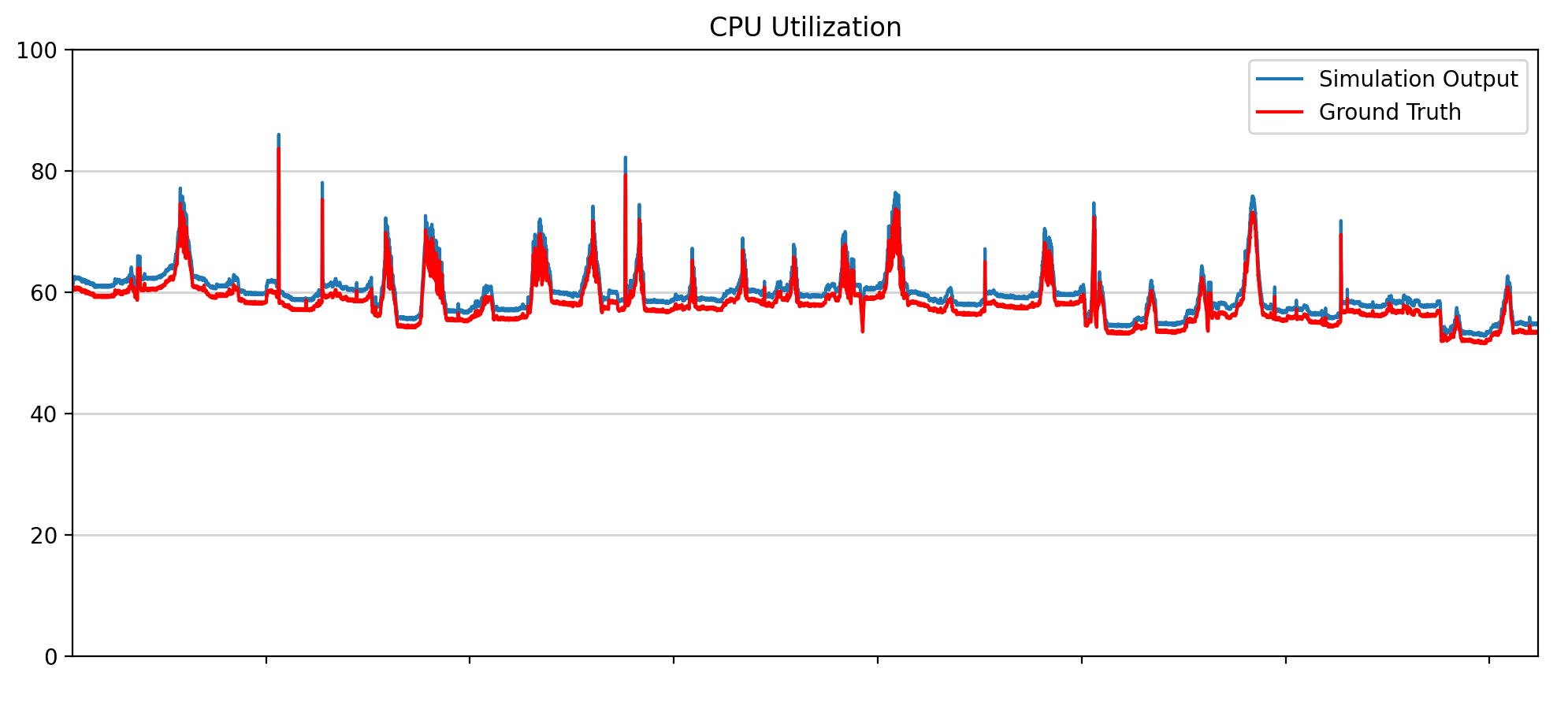}
    \caption{Validation of our simulator (y axis in \%).}
    \label{fig:imple:sim_fidelity}
\end{figure}

\textbf{Simulator Warm-Up}: One challenge in simulating our workloads is reconstructing the exact state of the  scheduler at the start of our trace, since not all system state was consistently saved. We therefore warm up our simulation by collecting all VMs that are live at the start of our trace, replay their start events in order, and then let the simulation run for another 2 simulated days to reach steady state. Warm up addresses left censorship of the trace and ensures that we run our simulations based on a snapshot that is representative of production behavior \emph{before} lifetime-based scheduling is enabled. As we will discuss in Appendix~\ref{sec:appendix:ablations}, this state is representative of a production setup where lifetime-based scheduling is enabled while the system is running. However, this methodology can reduce the impact of lifetime aware scheduling, since it takes a long time for all VMs that were placed without lifetime-based scheduling to exit.

\textbf{Computing Stranding}: While we directly measure the impact of our approach on empty hosts in simulation, stranding is computed via a separate pipeline (Section~\ref{sec:background:constraints}). We thus integrate our production stranding computation pipeline with our simulator, to compute stranding at a given point in time.

\textbf{Validation}: We validate our simulator by comparing its behavior to production numbers during the same time interval. We found it to be highly accurate during our validation studies (Figure~\ref{fig:imple:sim_fidelity}). For example, the  CPU utilization across the cluster was on average within $1.59$\% of the ground truth (with a standard deviation of $0.23$\%). We also found that the fragmentation numbers from our production experiments (Section \ref{sec:eval:prod}) closely matched the simulated numbers, further supporting the validity of the simulator.

Note that the fixed offset in Figure~\ref{fig:imple:sim_fidelity} does not imply that we can easily close this gap. The offset largely stems from dynamically invested/divested capacity that can result in small fluctuations of the pool size. The offset is not always constant and can be positive or negative.

\section{Additional Evaluation Results}
\label{sec:appendix:detailed-eval}

We now present additional data to further support our results in Section~\ref{sec:eval} and provide more details.

\subsection{Accuracy-Performance Trade-off}
\label{sec:appendix:em-tradeoff}

\begin{figure}
    \centering
    \includegraphics[width=\linewidth]{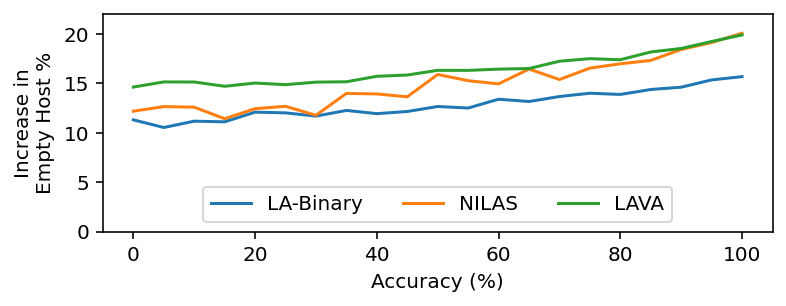}
    \caption{Performance at different levels of prediction accuracy.}
    \label{fig:eval:accuracy_em_tradeoff}
\end{figure}

\begin{figure}
    \centering
    \includegraphics[width=0.98\linewidth]{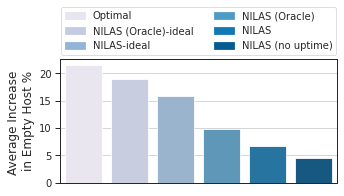}
    \caption{NILAS using oracular lifetime run at ideal setting (cold start and highest priority) achieves near-optimal performance. NILAS also consistently outperforms the version of NILAS that does not use repredication or uptime. The reported numbers are averaged across running traces from 24 C2 pools in simulation.}
    \label{fig:eval:las_accuracy}
\end{figure}

We plot the trade-off curve between model accuracy and performance (Figure~\ref{fig:eval:accuracy_em_tradeoff}). We start from our oracular predictions and randomly categorize each VM into one of two buckets – correctly predicted or incorrectly predicted.  The probability of these two cases is governed by the accuracy on the x axis. Then, we apply a different Gaussian error distribution to the label ($\sigma$ = 0.001 for correctly predicted VMs and $\sigma$ = 3 for incorrectly predicted VMs, in the Log10 domain). To be more representative of the actual model's behavior, we also cap lifetimes to [0, 14 days].

We see that our improvements persist with different accuracy values, and that LAVA is better than NILAS at tolerating high rates of mispredictions, as expected.

\subsection{Ablations \& Theoretical Limit}
\label{sec:appendix:ablations}

We next conduct a focused study of NILAS to understand how far it is from the theoretical limit, and what factors contribute to the remaining gap.

Since all server host hardware is the same within each pool we tested, we can compute the optimal percentage of empty hosts based on the total fraction of un-reserved resources (the lower of two resource dimensions: CPU and memory) aggregated across all hosts in that pool. This optimal value sets an absolute upper bound on the total number of hosts that can be made empty under a given pool size and load.

We then test NILAS under an \emph{ideal} simulation setting to find out the best performance it can attain. In this ideal setting, NILAS uses oracular lifetime and is put as the highest-ranked scoring function without the simulator's warm-up phase (Appendix~\ref{sec:appendix:simulator}). Skipping warm-up amplifies the impact of NILAS since it allows NILAS to schedule VMs onto an empty cluster leveraging lifetime information throughout the whole trace, without the residual impact of VMs that were already scheduled without lifetime-awareness during the warm-up phase.

Figure~\ref{fig:eval:las_accuracy} shows that ideal runs of NILAS get very close to the optimal result --  the greedy approach taken by the NILAS algorithm is nearly optimal. However this ideal setting is unlikely to be reproduced in our production setting, where gradual roll-out (mimicked by the warm-up phase in simulation) is necessary and there are more critical scheduling criteria that rank above NILAS. These factors combined contribute to the lower performance achieved by NILAS with oracular lifetime but a non-\emph{ideal} setting. Mispredictions from the model further bring down the amount of savings by NILAS. Additionally, we observe that not using repredictions cause more mistakes, which leads to significantly reduced performance and demonstrates the importance of our reprediction-based approach.

\begin{figure}
    \centering
    \includegraphics[width=\linewidth]{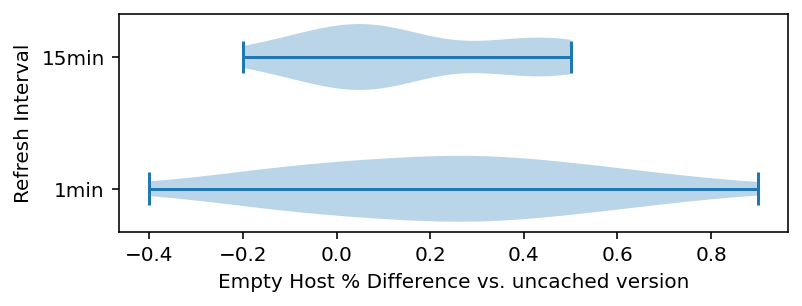}
    \caption{Effect of caching predictions, across 22 pools (in simulation). Note that repredictions are still performed when a VM is added or removed from a host, irrespective of the refresh interval.}
    \label{fig:eval:caching}
\end{figure}

\subsection{Ablation: Caching Predictions}
\label{sec:appendix:caching}

As stated in Section~\ref{sec:implementation}, we found that repredicting every VM on each considered host becomes a bottleneck in some very large pools. We therefore introduced a caching approach. Here, we provide an ablation study to show that caching predictions and repredicting them at a coarser granularity does not adversely affect the performance of NILAS. Figure~\ref{fig:eval:caching} shows these results: We compare simulations of our approach with caching against the baseline without cached predictions. We show two different caching intervals: 1min and 15min. We note that this interval only affects hosts that see no VM changes for extended periods of time; as mentioned in Section~\ref{sec:implementation}, when a VM is added to or removed from a host, the predictions on this machine are updated irrespective of the refresh interval. On average, we see a small improvement – we hypothesize that this could be due to caching addressing the ``dip'' in Figure~\ref{fig:eval:uptime_accuracy}.

\section{LARS Algorithm Details}
\label{sec:appendix:lars}

This section provides additional details about  defragmentation and improving it with Lifetime Aware ReScheduling (LARS).

\begin{algorithm}
\caption{LARS Algorithm}
\KwIn{A set of candidate hosts $C$ for eviction}

\For{each candidate $c \in C$}{
    $v_{sorted}$ = Sort the existing VMs $V_c$ on $c$, based on their predicted remaining lifetime, in descending order

    Send all VMs for eviction approval

    \If{not all VMs are approved for eviction}{
        continue}
    \Else{Stop scheduling new VMs on $c$
    
        \For{$v \in v_{sorted}$}{
          Allocate new VM of the same shape as $v$
          
          Live migrate $v$ to this new VM
          
          Once finished, remove $v$ from $c$
        }
    }
}
\end{algorithm}

Defragmentation is triggered when the number of empty hosts in a particular pool drops below a particular threshold.  The defragmenter picks a set of candidate hosts $C$ based on factors such as their current occupancy.  It then evacuates the VMs on each host, live migrating~\cite{ruprecht2018vm} them by copying them to another host. The scheduler selects the target host, using the same algorithm it uses for new VMs, but with the current VM state (e.g., lifetime prediction, resources, etc.). 

We show the pseudo-code for this operation below. Once evacuation of a particular host begins, the defragmenter first confirms with a higher-level system that live migration of these VMs is consistent with system-level objectives. Once approved, the Borg stops scheduling new VMs on this host. (In the absence of live migration, this host would become empty as a function of the VMs' exit times).  On each host, live migration chooses VMs to reschedule one at a time. Live migration occurs on multiple hosts concurrently, up to a configurable limit.

The first step of live migration is to choose a target host. Live migration uses the same scheduling algorithm as Borg uses when initial scheduling a VM (e.g., NILAS, LAVA, or the original waste-minimization scheduler). Since the VMs have already been running for a period of time, live migration with NILAS and LAVA may already lead to improved placement because they take repredicted VM lifetimes into account.  Once Borg allocates memory for the migrating VM on the new host, it migrates the original VM, and once migration has completed, the old instance is deleted. Once Borg migrates all the VMs, the host is empty and can be updated (in the case of system maintenance), put in low power mode, or divested/powered down.

The modification that LARS makes to this algorithm is that it performs the VM live migrations in the order of the predicted remaining lifetime. This optimization gives short-lived time to naturally exit while other VMs migrate. Each such VM saves one live migration.

\section{NILAS \& LAVA Algorithm Details}

We provide pseudo-code for the NILAS and LAVA algorithm, to further formalize and clarify its behavior.

\begin{algorithm}
\caption{NILAS Scheduling Algorithm}
\KwIn{A new VM $v$}
\KwOut{Schedule $v$ to a host}

Find host $h$ such that $v$ does not exceed the largest exit time of VMs on $h$

\If{No such $h$ exists}{
Schedule $v$ on host $h$ (with capacity to accommodate $v$) such that largest exit time of VMs on $h$ is changed by least amount upon scheduling $v$ on $h$}

\If{No such $h$ exists}{
    Indicate $v$ fails to be scheduled}
\end{algorithm}

\newpage

\begin{algorithm}
\caption{LAVA Scheduling Algorithm}
\KwIn{A new VM $v$}
\KwOut{Schedule $v$ to a host or give scheduling failure}

\If{there exists a recycling host $h$ where $h.LC > v.LC$ and $v$ can fit on $h$}{
    Let $C$ be the set of recycling hosts $h'$ where $h'.LC$ is closest to $v.LC$
}
\ElseIf{there exists a matching host $h$ where $h.LC = v.LC$ and $v$ can fit on $h$}{
    Let $C$ be the set of matching hosts
}
\ElseIf{there exists a non-empty host $h$ where $v$ can fit on $h$}{
    Let $C$ be the set of non-empty hosts
}
\ElseIf{there exists an empty host $h$ where $v$ can fit on $h$}{
    Let $C$ be the set of empty hosts
}
\Else{
    Indicate that $v$ fails to be scheduled
    \Return
}
Schedule $v$ to the host $h^* \in C$ selected by $NILAS$ 
\end{algorithm}

\end{document}